\newif\iflong\longfalse
\newif\ifmacro\macrofalse
\documentclass[conference]{IEEEtran}
\usepackage[utf8]{inputenc}

\usepackage{amssymb,amsmath,amsthm,color}

\usepackage{xspace}

\usepackage{stmaryrd}

\usepackage{enumitem}

\theoremstyle{definition}
\newtheorem{definition}{Definition}
\newtheorem{lemma}[definition]{Lemma}
\newtheorem{prop}[definition]{Proposition}
\newtheorem{proposition}[definition]{Proposition}

\newtheorem{example}[definition]{Example}
\newtheorem{theorem}[definition]{Theorem}
\theoremstyle{remark}
\newtheorem{remark}[definition]{Remark}

\usepackage{mathpartir}

\newcommand{\ofnames}[1]{\ensuremath{\mathrm{fn_o}(#1)}}

\mathcode`\!="4021 
\mathcode`\.="602E  
\mathcode`\|="326A 

\newcommand{\rref}[1]{(\ref{#1})}

\newcommand{\ttag}{\ensuremath{\eta}}
\newcommand{\typs}[2]{\ensuremath{#1\vdash_{\mathrm{s}}#2}}
\newcommand{\typsR}[3]{\ensuremath{#1;#2\vdash_{\mathrm{s}}#3}}
\newcommand{\typr}[2]{\ensuremath{#1 \vdash_{\mathrm{wb}}#2}}

\newcommand{\typedOKp}[2]{\ensuremath{#1\vDash #2}}
\newcommand{\typps}[2]{\ensuremath{#1\vDash_{\mathrm{s}}#2}}
\newcommand{\typpsR}[3]{\ensuremath{#1;#2\vDash_{\mathrm{s}}#3}}
\newcommand{\typpr}[2]{\ensuremath{#1 \vDash_{\mathrm{wb}}#2}}

\newcommand{\OR}{\ensuremath{~\big|~}}

\newcommand{\many}[1]{\ensuremath{\widetilde{#1}}}
\newcommand{\inpi}[2]{#1(#2)}
\newcommand{\earlyinpi}[2]{#1\langle#2\rangle}
\newcommand{\outpi}[2]{\ensuremath{\overline{#1}\langle #2\rangle}}

\newcommand{\out}[1]{\ensuremath{\overline{#1}}}
\newcommand{\new}{\ensuremath{\boldsymbol{\nu}}}
\newcommand{\respi}[1]{\ensuremath{(\new #1)}}
\newcommand{\nil}{\ensuremath{\boldsymbol{0}}}

\newcommand{\rconf}[2]{\ensuremath{[#1 ; #2]}}
\newcommand{\hole}{\contexthole}

\newcommand{\psapp}[2]{\ensuremath{#1_{#2}}} 

\makeatletter
\newcommand{\xMapsto}[2][]{\mapstochar \ext@arrow 0359\Rightarrowfill@{#1}{#2}}
\newcommand{\xmapsto}[2][]{\mapstochar \ext@arrow 0359\rightarrowfill@{#1}{#2}}
\makeatother

\newcommand{\strans}[1]{\ensuremath{\ired{#1}}}
\newcommand{\sttrans}[4]{\ensuremath{\typps{#1}{#2 \ired{#3} #4}}}
\newcommand{\sttransR}[5]{\ensuremath{\typpsR{#1}{#2}{#3 \ired{#4} #5}}}
\newcommand{\wttrans}[4]{\ensuremath{\typps{#1}{#2 \Ired{#3} #4}}}

\newcommand{\wtrans}[1]{\Ired{#1}}
\newcommand{\red}{\strans{}}
\newcommand{\wred}{\wtrans{}}
\newcommand{\redd}{\strans{}_{\textrm{d}}}
\newcommand{\wredd}{\wtrans{}_{\mathrm{d}}}

\newcommand{\wbarb}[1]{\ensuremath{\Downarrow_{#1}}}

\newcommand{\R}{\ensuremath{\mathcal{R}}}
\newcommand{\RR}{\ensuremath{\mathrel{\mathcal{R}}}}
\newcommand{\biswb}{\ensuremath{\approx_{\rm{wb}}}}
\newcommand{\bcong}[1]{\ensuremath{\cong^{#1}}}
\newcommand{\beq}[1]{\ensuremath{\simeq^{#1}}}
\newcommand{\wAbis}{\ensuremath{\approx}} 
\newcommand{\wbiss}[1]{\ensuremath{\approx_{\rm{s}}^{#1}}}
\newcommand{\wbissn}[2]{\ensuremath{\approx_{\rm{s}}^{#1,#2}}}
\newcommand{\wbRS}[2]{\ensuremath{\approx_{\rm{s}}^{#1;#2}}}

\newcommand{\bbis}[1]{\ensuremath{\mathrel{\overset{\mbox{\large .}}{\approx}^{#1}}}}
\newcommand{\bisup}{\ensuremath{{\wredd\RR^{\rm{C}}}}}
\newcommand{\fnames}[1]{\ensuremath{\mathrm{fn}(#1)}}
\newcommand{\bnames}[1]{\ensuremath{\mathrm{bn}(#1)}}

\newcommand{\defi}{\ensuremath{\stackrel{def}{=}}}

\newcommand{\set}[1]{\ensuremath{\left\{#1\right\}}}

\newcommand{\Ima}[1]{\ensuremath{\mathrm{codom}(#1)}}
\newcommand\codom\Ima

\newcommand{\sz}[1]{{|}#1{|}}

\newcommand{\engue}[1]{\ifmacro{\color{cyan}\bfseries{\huge\ensuremath{\star}} {\footnotesize#1}}\fi}
\definecolor{re}{rgb}{0.0,.40,0.4}
\definecolor{funcolor}{rgb}{.528,.266,.595}  
\definecolor{impcolor}{rgb}{.804,.4,.114}   

\newcommand{\rfaaD}[2]{\ensuremath{\mathtt{faa2}_\LLL\langle #1\rangle(#2)}}
\newcommand{\rswapD}[2]{\ensuremath{\mathtt{sw2}_\LLL\langle #1\rangle(#2)}}

\newcommand{\LLL}{\ell}
\newcommand{\rreadL}[1]{\ensuremath{\mathtt{re}_\LLL(#1)}}
\newcommand{\rwriteL}[1]{\ensuremath{\mathtt{wr}_\LLL\langle #1\rangle}}
\newcommand{\rfaaL}[2]{\ensuremath{\mathtt{faa}_\LLL\langle #1\rangle(#2)}}
\newcommand{\rswapL}[2]{\ensuremath{\mathtt{sw}_\LLL\langle #1\rangle(#2)}}

\usepackage{davidemacros}
\def\ds#1{\ifmacro\vskip.2cm\noindent{\em #1}%
  \marginpar{{{\bf{DS}}}}\vskip.2cm\fi}
\def\daniel#1{\ifmacro\vskip.2cm\noindent{\em #1}%
  \marginpar{{{\bf{DH}}}}\vskip.2cm\fi}

\author{\IEEEauthorblockN{Daniel Hirschkoff}
  \IEEEauthorblockA{ENS de Lyon}
  \and \IEEEauthorblockN{Enguerrand Prebet
  }
  \IEEEauthorblockA{ENS de Lyon}
  \and \IEEEauthorblockN{Davide Sangiorgi
}
  \IEEEauthorblockA{Università di Bologna and INRIA}
}

\title{On sequentiality and well-bracketing in the $\pi$-calculus}

\IEEEoverridecommandlockouts
\IEEEpubid{\makebox[\columnwidth]{
    978-1-6654-4895-6/21/\$31.00~\copyright2021 IEEE \hfill} \hspace{\columnsep}\makebox[\columnwidth]{ }}

\begin{document}

\maketitle

\begin{abstract}
The  $\pi$-calculus  is used as 
 a model for programming languages. 
Its contexts exhibit arbitrary concurrency, making them very discriminating. 
This may prevent validating desirable behavioural equivalences in cases  
when more disciplined contexts are expected. 

In this paper we focus on two such common  disciplines: sequentiality, meaning that at
  any time there is a single thread of computation,   and  well-bracketing, 
meaning that calls to external services obey a stack-like discipline.
We formalise the disciplines by means of type systems. 
The main focus of the paper is  
on  studying the consequence of the disciplines on behavioural
  equivalence.
We define and study labelled bisimilarities for sequentiality  and well-bracketing. These
relations are coarser than ordinary bisimilarity. 
 We prove that they are sound for the respective (contextual)  barbed equivalence, and also
  complete 
 under a certain technical condition. 

We show the usefulness of our techniques on a number of examples, that have mainly to do
with the representation of functions and store. 



\end{abstract}

\section{Introduction}\label{s:intro}

The $\pi$-calculus has been advocated as a model to  give 
semantics to, and reason about, various forms of programming languages, including those 
 with higher-order
features. Strengths of the $\pi$-calculus are its rich algebraic
theory and its wide spectrum of  proof techniques. 
Concurrency is at the heart of the $\pi$-calculus:
computation is interaction between concurrent processes. 
The  operators of the calculus are simple  (parallelism, input, output,
restriction being the main ones) and unconstrained. 
This yields
 an amazing expressive power~--- the calculus
 can model a variety of
programming idioms~\cite{SanWal}. However, 
this also   makes 
the contexts of the calculus very discriminating;  
as a consequence,  
behavioural equivalences,
which are supposed to be preserved by all
the contexts of the calculus,  
 are rather demanding relations.

Higher-level languages may be syntactically quite different  from a language for pure
concurrency such as the $\pi$-calculus.  
For instance, the
paradigmatic higher-order programming language,
 the $\lambda$-calculus,  is a pure calculus of functions  
and, in  both its 
call-by-name and call-by-value variants,  is  
 sequential~--- it is even deterministic.   
A variety of extensions of it have been considered; examples of additional features are
references, control operators, non-determinism,  (constrained) forms of concurrency. 
The specific  set of syntactic features chosen for  the language determines the ways in which the
contexts of the language may interact with the terms.  In any case, the patterns
of interaction are usually more disciplined  than those that arise in 
$\pi$-calculus  representations of those terms. 
Thus 
there are $\lambda$-terms that are indistinguishable within
the (pure) $\lambda$-calculus  whose $\pi$-calculus images 
can be separated by appropriate $\pi$-contexts.



A  well-known way of imposing a discipline to the $\pi$-calculus is to equip it with a  type system. 
Such systems are intended to capture communication patterns that occur
frequently when programming in the $\pi$-calculus. A  number of type
systems have been considered: e.g.,  capability types (formalising
 the  intended I/O usage of names that are exchanged among
processes),  linearity (formalising the property that certain
names may be used at most  once),  session types
(formalising the communication protocols in the dialogues between two
or more processes), and so
on~\cite{DBLP:journals/mscs/PierceS96,DBLP:journals/toplas/KobayashiPT99,DBLP:conf/esop/HondaVK98,AnconaBB0CDGGGH16}. 
Type systems have also been
designed to capture specific properties of processes, such as
termination, deadlock-freedom, lock-freedom \cite{DBLP:journals/toplas/Kobayashi98,DBLP:journals/iandc/Kobayashi02,DBLP:journals/iandc/DengS06,DBLP:journals/iandc/YoshidaBH04,DBLP:journals/toplas/KobayashiS10}.
Types impose constraints on the set of legal contexts in which
well-typed terms are supposed to be used; this can make behavioural
equivalences usefully coarser. 

A further  step is then  to tune the proof techniques of the $\pi$-calculus
to such type systems, so to be able to actually prove the behavioural equalities
that only hold in presence of types.  Typically this is 
investigated in the coinductive setting of bisimilarity, and 
achieved by
 refining and/or modifying   the standard
bisimilarity clauses so to take the usage of types  into
account. 
The resulting bisimilarity should be sound with respect to
contextually-defined forms of bisimilarity such as \emph{barbed equivalence} (or  \emph{congruence});
ideally, it should also be complete.

In barbed  \equivalence, the bisimulation game is played only on internal actions, and
certain success signals, the barbs, are used to monitor the computation.
In the standard barbed \equivalence, an arbitrary  context may be added, once (at the beginning), on
top of the tested processes. In \emph{reduction-closed} barbed \equivalence \cite{HoYo95,SanWal}, the context may be
dynamically updated, by adding 
further components during the computation. 
Reduction-closed barbed \equivalence usually allows simpler proofs of completeness, and 
 does not require 
any hypothesis of image-finiteness on the state space of the
tested processes. 
In contrast, standard barbed \equivalence is more robust~--- reduction-closed barbed \equivalence may
sometimes  be over-discriminating~\cite{SaWa01}.

In this paper we focus on the $\pi$-calculus representation of
\emph{sequentiality}  and \emph{well-bracketing}.
`Sequentiality' intuitively indicates the existence of a 
 single thread of computation.
'Well-bracketing' is  a  terminology
 borrowed from
 game semantics, and
 used to refer to 
 a language 
without control operators, in 
which the call-return interaction behaviour
 between a term and its context follows a stack discipline.
Our main objectives are to define bisimilarity-based proof techniques for type systems 
in the $\pi$-calculus 
that  formalise the sequentiality  and well-bracketing notions.
We actually  work with the \emph{asynchronous} $\pi$-calculus, \Api, as this is
the calculus that is usually adopted in the literature for modelling
higher-order languages.

In    \Api, 
 sequentiality is the property that, at any time, 
at most one   process  \emph{is active},  or \emph{carries the thread}; that is, 
the process has the control on
the computation and  decides what the next computation step can be.
In other words, we  
 never find two sub-components of a system  both
of which contain an \emph{interaction redex} 
(a pair of an input and an output processes at
the same name).

In the (standard) encodings of the $\lambda$-calculus \cite{DBLP:journals/mscs/Milner92,DBLP:journals/iandc/Sangiorgi94}, 
a process is active, i.e., it carries the thread,  when it contains an unguarded output
particle.
Indeed, 
the $\pi$-calculus terms obtained from the encodings give rise to computations  
in which,  syntactically,  at any time there is at most one unguarded output particle.
 An input process that consumes that output will in
turn become active.

Our type system is  more general, in that we allow 
also input processes to carry the thread. 
The type system specifies whether a name may carry  the thread in output
or in input; we call these names \emph{output-controlled} and \emph{input-controlled}.    
While the output-controlled are the most important ones (for instance, they play a central
role in the modelling of functions), input-controlled
names may be 
 useful too, for instance, 
in the representation of references or locks. 
A reference $\ell$ that contains the value $n$ is represented in  \Api
 by an output particle $\outpi \ell n$;  and a process
accessing the reference will do so by performing  an input at $\ell$. Thus
an input at 
 $\ell$  indicates ownership of the current 
computation thread.

As remarked above, sequentiality implies  absence of parallel
computation threads.
Sequentiality however does not exclude non-determinism. An output particle
$\outpi ab$ that owns the thread may have the possibility of interacting
with different input processes at $a$ (and symmetrically 
for input processes 
owning the thread). 
 Indeed we also admit internal
non-determinism (i.e, processes such as $\tau.P + \tau.Q$ that
 may chose to reduce either to $P$
or to $Q$   without
interactions with the environment), both in active and in inactive processes. 

The type system for well-bracketing is a refinement of that for sequentiality, in which a
stack of \emph{continuation names} keeps track of the structure of calls and returns among
the processes.   These stacks are similar to those used in the implementation of compilers
for  languages (or fragments of  languages)  adopting  well-bracketing, or used in
well-bracketed forms of game semantics. 


Finding proof techniques to reason about sequentiality and well-bracketing
 presents a number of caveats, that have mainly
to do with the soundness and completeness of the resulting
bisimilarity with respect to barbed \equivalence.  We briefly discuss
below a couple of  issues  concerning completeness. 

In the proof of completeness   one has to  show  that the contexts  of
the language are at least as discriminating as the labelled bisimilarity. 
In  standard proofs, one defines special  contexts that interact
with the tested processes and, at the same time, emit certain signals
to the outside so to provide information on the kind  of interactions
that have occurred with the processes. Such behaviour of the testing
contexts  is however inherently concurrent~--- the context has to interact with 
the tested processes
and, at the same time, 
emit  signals to the outside~---
and is therefore liable to break  the typing discipline for
sequentiality (and hence also well-bracketing).

Further problems arise in  proofs about  reduction-closed barbed \equivalence.
The reason why completeness proofs 
for reduction-closed barbed \equivalence
may be simpler than with  standard barbed \equivalence 
is that the testing context may be incrementally adjusted, after every
interaction step  with the tested processes. 
This however requires the existence of  special components in the
 contexts  to handle the fresh names generated by
 the tested processes. Specifically, the task of these components is to ensure that 
 new pieces of contexts,   added later, 
will be able to access  such fresh names. Again, these components represent parallel
 threads, and break the sequentiality and well-bracketing
 disciplines.  For this reason in the paper we cannot appeal to
reduction-closed forms of barbed \equivalence, remaining within the standard
 notions and therefore requiring an image-finiteness condition.

In the case of well-bracketing the problems above are enhanced by the presence of
continuation names. These names 
 are \emph{linear}~\cite{DBLP:journals/toplas/KobayashiPT99} (they may only be used once),
 \emph{input receptive}  \cite{DBLP:journals/tcs/Sangiorgi99} 
(the  input-end of the name should always be  available), and output-controlled. 
This places further constraints on the use of such names within
contexts that test the processes. 

For the above reasons, the completeness proofs for sequentiality and well-bracketing
present significant technical differences, both between them and from completeness proofs
in the literature. 

In the paper we propose labelled bisimilarities that allow us to reason about processes
following the sequentiality or well-bracketing  disciplines. We prove that the
bisimilarities are sound with respect to barbed equivalence. We also establish
completeness, on processes with only output-controlled names.  We do not know whether
completeness holds in the general case, with also input-controlled names. 
\ds{so we say somewhere why the case of input-controlled names is
  problematic? }
\daniel{the following sentence is just after the proof sketch for
  completeness-seq:\\
  \textit{The reasoning sketched above does not apply if input-controlled names
are allowed, intuitively because this introduces processes that
exhibit no barbs while controlling the thread, such as, e.g., $u.\out x$.
}}
We also study some refinements of the bisimilarities: one is obtained by injecting ideas
from bisimilarities for calculi with references \cite{DBLP:conf/concur/HirschkoffPS20};  other refinements are
forms of  `up-to techniques'.  
We illustrate applications of our techniques on a number of examples, most of which have
to do with the representation of functions and references. Usually the examples are about equalities
that only hold under the sequentiality or well-bracketing disciplines;
other examples show that  sequentiality and well-bracketing
may make   equalities  simpler to prove because there
are fewer observables to take into account.  


\emph{Paper outline.} We introduce some background in
Section~\ref{s:api}. We study sequentiality in Section~\ref{s:seq},
and well-bracketing in Section~\ref{s:wb}: in each case, we present
our type system, define an appropriate notion of bisimilarity, and
show some examples or laws that we can derive. Related  and future works
are discussed in Section~\ref{s:ccl}.
For lack of space, some technical definitions and proofs are given in~\cite{HPS:lics21:long}.

\section{Background: the (asynchronous) $\pi$-calculus}\label{s:api}
We recall here the standard  syntax of the asynchronous $\pi$-calculus, \Api, from~\cite{DBLP:journals/tcs/AmadioCS98}:
$$
\begin{array}{rcl}
     P,Q& ::=& \outpi{a}{\many b} \OR !\inpi{a}{\many b}.P \OR P|Q \OR \respi{a}P \OR G
		\\[.1em]
  G,G'& ::=& \nil \OR \inpi{a}{\many b}.P \OR \tau.P \OR [a=b]G \OR G+G'
\end{array}
$$
	Names are ranged over by $a,b,...$. 
In prefixes $\outpi{a}{\many b}$ and $ \inpi{a}{\many b}.P $, name $a$ is the 
\emph{subject} and $\til b$ are the \emph{objects}. 
We use a tilde, like in
        \many b, for (possibly empty) tuples of names; similarly 
        $\respi{\many{a}} P$ stands for a sequence of restrictions. 
 As usual,  we write $a.P$ and $\out a$ when the object of a prefix  is the empty tuple.
%
 We use $\sum_{i\in I} G_i$ (resp. $\prod_{i\in I} P_i$) for $G_{i_1}
 + \dots + G_{i_n}$ (resp. $P_{i_1} | \dots | P_{i_n}$) where $I =
 \set{i_1,\dots,i_n}$.
 We write $P\sub a b$ for the result of replacing name $b$ with $a$ in
 $P$ in a capture-avoiding way.
 Contexts, $C$, are processes containing a single occurrence of a
 special constant, the
 hole (written $\hole$).
The {\em static\/} contexts, ranged over by $E$,
have the  form $\res{\many a}(P | \hole)$.
 In examples, for readability we sometimes use  basic data values such as integers
 and booleans.
	The definition of structural congruence, written $\equiv$,  
and of the strong and weak labelled
        transitions, written
        $\strans{\mu}$, $\wred$, and $\wtrans{\hat\mu}$,  are standard and are given in~\cite{HPS:lics21:long}.
We note $\fnames{P}$ (resp. $\fnames{\mu}$) the set of free names of $P$ (resp. $\mu$).
We sometimes abbreviate reductions  $P \arr\tau P' $ as $P \longrightarrow P'$.

The calculi in the paper will be typed. For simplicity we define our type systems as
refinements 
of the  most basic type system for $\pi$-calculus, namely 
 Milner's {\em sorting} \cite{Mil91}, in which 
names  are partitioned into a collection of {\em types} (or sorts), and  
a sorting  function maps types onto types. If a  name type
$S$  is mapped onto a type  $T$, this means that  names in $S$ may
 only carry names in $T$.
We  assume that there is a sorting system under which all
processes we manipulate are well-typed.  
We write $\typedOK \Delta  P$ when process $P$ is well-typed under
$\Delta$,
and similarly for other objects, such as contexts.

The reference behavioural equivalence  for us will be the context-closure of barbed
bisimulation. We focus on barbed equivalence (as opposed to barbed congruence) because it
is simpler (notably, we do not need to consider issues of closure of the labelled bisimulations under name
substitutions).   
The definition of barbed bisimulation  uses  the
reduction
relation $\Longrightarrow $ along with an observation
predicate $\Dwa_a$ for each name $a$, which detects  the
possibility of performing  an output to the external  environment along
$a$.  Moreover, since we work in a typed setting,  such an output should be allowed
by the typing of the tested processes. 
 Thus, we write  $\typedOKp \Delta 
{P \Dwa_a}$  if $\Delta$ is a typing for $P$ (i.e.,  $\typedOK \Delta P$ holds), 
there is an output $\mu$ with  subject $a$
 s.t.\ 
$P \Arr\mu P'$,  and such a transition is observable under the typing $\Delta$.
The meaning of 'observable under a typing' will depend on the specific type system
adopted; in the case of the plain sorting, all transitions are observable.

Having typed processes, in the definition of barbed equivalence we may only test 
processes with 
 contexts
that respect the typing of the processes. 

\begin{definition}
\label{d:stCtT}
  $\qct$ is a \emph{$\conTT \Gamma {\Delta }$  context}  if 
$\typedOK{ \Gamma }{\qct}$ holds, using the typing for the processes 
plus the
 rule $
 	\inferrule{ }{\typedOK{ \Delta}{\contexthole}}
$ for the hole.
\end{definition} 

Similarly, $P$ is a \emph{$\Delta$-process} if 
$\typedOK \Delta  P$.  
We assume (as in usual Subject-Reduction properties for type systems)
that typing is invariant under reduction.
\iflong
This is the case for the type systems we present in this paper.
\fi
\begin{definition}[Barbed bisimulation, equivalence, and congruence]
\label{d:bb} {\em Barbed $ \Delta$-bisimulation}
is the largest symmetric relation $\bbis \Delta$ 
 on  $\Delta$-processes s.t.\ 
  $P \bbis \Delta  Q$ implies: 

\begin{enumerate}
\item   whenever $P  \longrightarrow  P'$ 
 then there exists $Q'$ such that
 $Q \Longrightarrow Q'$ and $P' \bbis \Delta  Q'$;

 \item   for each name  $a$,    
 $\typedOKp \Delta {P \Downarrow_a}$ iff     $ \typedOKp \Delta {Q \Downarrow_a}$.

\end{enumerate}
Two $\Delta$-processes $P$ and $Q$ are {\em barbed equivalent at $ \Delta$}, written
$P \beq \Delta   Q$, if  for each 
$\conTT \Gamma {\Delta }$ static  context $E$
it holds that  $E[ P] \bbis\Gamma   E[ Q]$.
\emph{Barbed congruence at $\Delta$},  $\bcong \Delta$, is defined in the same way but
employing all  $\conTT \Gamma {\Delta }$  contexts (rather than only the static ones). 
\end{definition}

Barbed equivalence in the plain (untyped)
 \Api,  $\beq{}$, 
can be proved to coincide with 
the ordinary
labelled early asynchronous 
bisimilarity, 
 on 
             image-finite 
processes,
 exploiting  the $n$-approximants of the  labelled
equivalences. 
 We recall
that 
the class of {\em image-finite processes} is the largest subset ${\cal I}$ of
 processes 
that is derivation  closed  and s.t.\
$P \in {\cal I}$ implies that,   for all $\mu$, 
the set 
$\{P'  \; \st\;   P \Arr{\mu } P'\}$, quotiented by   alpha 
conversion, is finite.
In the remainder of the paper, we omit the 
 adjectives `early'  and `asynchronous' in all bisimilarities. 
\iflong
Similarly, the relation we present below is usually called
\emph{asynchronous bisimulation}, due to the second clause for input
transitions; we simply call it
\emph{bisimulation}, and do so for the coinductive equivalences we consider in
the paper, which are all asynchronous.
       \fi

	\begin{definition}[Bisimulation]
\label{d:bisa}
		A relation $\R$ on processes is a \emph{bisimulation} if whenever $P\RR Q$
                and $P\strans{\mu}P'$,  then one of these
		two clauses holds: 
	\begin{enumerate}
\item
	there is $Q'$ such that $Q\wtrans{\hat\mu}Q'$ and $P'\RR Q'$; 
			\item $\mu = \earlyinpi{a}{\many{b}}$ and there is $Q'$ such that $Q|\outpi{a}{\many{b}}\wtrans{}Q'$ and $P'\RR Q'$.
                        \end{enumerate} 
Moreover the converse holds too,  on  the transitions from  $Q$.
		\emph{Bisimilarity}, 
$\wAbis$, is 
 the largest
                bisimulation.
	\end{definition}

\begin{theorem}[\cite{DBLP:journals/tcs/AmadioCS98}]
  \label{t:bc}
On image-finite processes, relations $ \beq{}$ and $\wAbis$ coincide. 
\end{theorem} 


	\section{Sequentiality}\label{s:seq}

In this section we study sequentiality. We first formalise it by  means of a type system,
and then we examine its impact on behavioural equivalence.

\subsection{Type system}
As mentioned in Section~\ref{s:intro},  intuitively, sequentiality ensures us that at any time
  at most one interaction can occur in a system; i.e., there is a single computation
  thread.  
A process that holds the thread decides what the next interaction can
  be.  It does so by offering a single particle (input or output) 
that controls the thread. The process may offer multiple particles, but only one of them
may control the thread. The control on the thread attached to 
a particle is determined by the
subject name of that particle. 
A given name may exercise the control on the thread either in   output or in
input; in the former case we say that the name is \emph{output-controlled},
  in the latter case the name is     \emph{input-controlled}.
For instance, suppose that
$x,y,z$ are  output-controlled and   $u,v$  are  
input-controlled.
Then the following process correctly manages the thread and will  indeed be  typable  in our
type system: 
\[ 
P \defi u.  (\outC x | y.\out{x}) | z. \outC y | \outC v    
\] 
The initial particles in $P$ are  $u, z,  \outC v$;  however only $u$ controls the thread, 
as $z$ is  output-controlled and $v$ is input-controlled.  When the input at $u$ is consumed, the
new particles $  \outC x, y$ are available, where $  \outC x$ now controls the thread, as
both names $x,y$ are output-controlled. 
An external  process that consumes the particle $\outC x$ will acquire the control over the thread. 
For instance,  a process such as  $Q \defi  \outC u | \inpC x . Q' $  initially does not hold the thread;
 in the parallel composition $P |Q$, after the two interactions at $u$ and $x$,
the control on the thread will be acquired by $Q'$:
\[ P | Q \longrightarrow \longrightarrow   (y.\out{x} | z. \outC y | \outC v)  | Q' \] 
Now $Q'$  will decide on the next interaction; for instance,
it may
offer an output at $y$ or $z$, or an input at $v$. It  may only offer  one of these,
though it may offer other particles that do not control the thread. 

\noindent 
{\bf Notation.} {\em
 In the remainder, 
$x,y,z$   range over   output-controlled
names,
 $u,v,w$  over input-controlled 
names; 
 we  recall  that $a,b,c$ range over the set of all names. 
}

The name used is therefore an indication of its type.
For instance,
  in $\respi xP$, $x$ is output-controlled, and can be only
  alpha-converted using another output-controlled name.

The type system for sequentiality is presented in Figure~\ref{f:typ:rule:seq1}. 
Judgements are of the form $\typs{\eta}{P}$, for $\eta \in \{0,1 \}$.
A judgement 
 $\typs{1}{P}$ indicates that  $P$ owns the thread, i.e., $P$ is  \emph{active}, and  
 $\typs{0}{P}$ otherwise, i.e., $P$ is  \emph{inactive}.

We recall that  we only present the additional typing constraints  given by sequentiality,
 assuming  the existence of a sorting under which all processes are well-typed
(thus the fully-fledged typing judgements would be the form 
$ \Delta ; \typs{\eta}{P}$, rather than $ \typs{\eta}{P}$).

Some remarks on the rules in Figure~\ref{f:typ:rule:seq1}:
a rule with a double conclusion  is an 
 abbreviation for more  rules with the same premises but separate
 conclusions.  The continuation of an input always owns  control on the thread; 
the input itself may or may not have the control (rules \trans{I-Act}
and  \trans{I-Ina}). A $\tau$-prefix is neutral w.r.t.\ the thread. 
The rule for parallel composition makes sure that the control on the thread is 
granted to only one of the components; in contrast, in the rule for sum, the control is
maintained for both summands. Operators $\nil$ and match cannot own the thread; this makes
sure that the thread control     is always exercised. 
	\begin{figure*}[t!]
	\begin{mathpar}
		\inferruleD{I-Act}{\typs{1}{P}} {\typs{1}{\inpi{u}{\many{a}}.P}}
		\and
		\inferruleD{I-Ina}{\typs{1}{P}} {\typs{0}{\inpi{x}{\many{a}}.P , \;  !\inpi{x}{\many{a}}.P}}
		\and
		\inferruleD{O-Act}{ }{\typs{1}{\outpi{x}{\many{a}}}}
		\and
		\inferruleD{O-Ina}{ }{\typs{0}{\outpi{u}{\many{a}}}}
		\and
		\inferruleD{Res}{\typs{\ttag}{P}}{\typs{\ttag}{\respi{a}P }}
		\and
		\inferruleD{Nil}{ }{\typs{0}{\nil}}
		\and
		\inferruleD{Par}{\typs{\ttag_1}{P} \and \typs{\ttag_2}{Q}}
		{\typs{\ttag_1+\ttag_2}{P|Q}}
                ~\ttag_1+\ttag_2\leq 1
		\and
		\inferruleD{Sum}{\typs{\ttag}{G_1} \and \typs{\ttag}{G_2}}{\typs{\ttag}{G_1 + G_2}}
		\and
		\inferruleD{Tau}{\typs{\ttag}{P}}{\typs{\ttag}{\tau.P}}
		\and
		\inferruleD{Mat}{\typs{0}{G}}{\typs{0}{[a=b]G}}
	\end{mathpar}
	\caption{The typing rules for sequentiality}
	\label{f:typ:rule:seq1}
	\end{figure*}

We present some behavioural properties that highlight the meaning of sequentiality. 
A reduction $P\arr\tau P'$ is an
        \emph{interaction} if it has been obtained from a communication between an
        input and an output (formally, its derivation in the LTS
        of~\cite{HPS:lics21:long} 
         uses rule \trans{AComm}). 
         \iflong
         A pair of   an unguarded input and an unguarded  output at the
        same name
form an  \emph{interaction redex}.  \fi
In a sequential system, one may not find two disjoint interactions.
\iflong interaction redexes. TOFIX\fi

	\begin{prop}
		Whenever $\typs{\ttag}{P}$, there exists no $P_1,P_2,
                \many{a}$ such that $P \equiv
                \respi{\many{a}}(P_1|P_2)$ with $P_1 \strans\tau P_1'$ and
                $P_2\strans\tau P_2'$, and  both these transitions are interactions.
	\end{prop}

An inactive process may not perform interactions. 

	\begin{prop}
If $\typs{0}{P}$, then 
\iflong
no interactions may occur in $P$; i.e., 
\fi
there is no $P'$ with $P \arr\tau P'$ and this transition is an interaction.  
\end{prop} 

An inactive process may however  perform $\tau$-reductions, notably to  resolve
internal choices. In other words such internal choices represent internal matters for a
process, orthogonal with respect to  the overall  interaction thread.  
The possibility for inactive processes to  accommodate  internal choices will be  important
in our completeness proof. 
However, an inactive process may 
only perform a finite number of $\tau$-reductions. 
A process $P$    \emph{is divergent} if it can perform an infinite sequence of
reductions, i.e., there  are $P_1, P_2, \ldots,$ with $P \longrightarrow P_1
\longrightarrow P_2 \ldots P_n   \longrightarrow  \ldots$.

	\begin{prop}
If $\typs{0}{P}$ then  $P$ is not divergent.
\end{prop} 

In contrast, an active process may be divergent, 
through sequences of reductions
containing infinitely many interactions. 

Sequentiality imposes constraints on the interactions that a `legal'  (i.e., well-typed) context
may undertake with a process.
For the definition of barbed bisimulation and equivalence we must therefore define the meaning of
observability. 
   The following definition of \emph{type-allowed transitions}
shows what such `legal'  interactions can be.

\begin{definition}[Type-allowed transitions]\label{d:tat}
  We write $\sttrans{\ttag}{P}{\mu}{P'}$ if $\typs{\ttag}{P}$,  and $P \strans{\mu} P'$,  and
  one of the following clauses  holds:
  \begin{enumerate}
  \item $\ttag = 0$
  \item $\mu = \tau$
  \item $\ttag = 1$ and $\mu = \earlyinpi{u}{\many{a}}$ for some $u,\many{a}$ or
    $\mu = \respi{\many{a}}\outpi{x}{\many{b}}$ for some
    $\many{a},x,\many{b}$. 
  \end{enumerate} 
\end{definition}
Clause (1)
says that all interactions between an inactive process and the context
are possible;
 this holds because the context is active and may therefore decide on the next
interaction with the process. 
Clause (2) says that internal reductions may always be performed. 
Clause (3) says that 
the only visible actions observable  in 
  active processes  are 
those carrying the thread; this holds because the observer is inactive, and it is therefore
up to the process to decide on the next interaction.

We now examine how typing evolves under legal  actions. We recall that $x$ stands
        for an output-controlled name. 

\begin{definition}
	We write $\rconf{\ttag}{P}\strans{\mu}\rconf{\ttag'}{P'}$ when
        $\sttrans{\ttag}{P}{\mu}{P'}$ and:
	\begin{enumerate}
		\item if $\mu = \earlyinpi{x}{\many{a}}$, then $\ttag' = 1$.
		\item if $\mu = \respi{\many{a}}\outpi{x}{\many{b}}$, then $\ttag' = 0$.
		\item otherwise $\ttag' = \ttag$.
	\end{enumerate} 
\end{definition}

\begin{theorem}[Subject Reduction]
\label{l:subred:seq1}
	If $\typs{\ttag}{P}$ and
	$\rconf{\ttag}{P}\strans{\mu}\rconf{\ttag'}{P'}$ then $\typs{\ttag'}{P'}$.
\end{theorem}
\iflong 
In the third clause above, in the case of an interaction, we
necessarily have $\ttag=1$.

\fi
Weak type-allowed transition are defined as expected, 
exploiting  the
 invariance of typing under reductions:  $\wttrans{\ttag}{P}{\mu}{P'}$  holds if  there are $P_0,P_1$ with 
$P \Longrightarrow P_0 $, $\sttrans{\ttag}{P_0}{\mu}{P_1}$ and
$P_1\Longrightarrow P'$.

\subsection{Behavioural equivalence}
To tune  Definition~\ref{d:bb} of  barbed bisimulation and equivalence to the setting of
sequentiality, 
we have to specify the meaning of  observables.
An observable  $\typps \eta  
{P \Dwa_a}$   holds if there are $P'$  and an output action $\mu$ such that 
 $\wttrans{\ttag}{P}{\mu}{P'}$ and the subject of $\mu$ is $a$.
Following Definition~\ref{d:stCtT}, in barbed equivalence,  the legal contexts are 
the  $\conTT \eta{\eta'}$  static contexts.
\iflong 
 To remind ourselves of the sequentiality constraint, 
we
\else
We 
\fi
 write 
barbed equivalence at $\eta$ as 
$\beq \eta$.
\iflong 
, and sometimes call it 
\emph{sequential barbed equivalence at $\ttag$}.
\fi
Thus  $P \beq \eta
 Q$  holds if
 $\typs{\ttag}{P,Q}$ and
$E[P] \bbis{\eta'}  E[Q]$, 
 for any $\eta'$ and 
 any $\conTT{\eta'} \eta $   static context $E$.



We are now ready to define the labelled bisimilarity to be used on
sequential processes, which is our main proof technique for barbed equivalence.
A \emph{typed  process relation} is a set of triplets $(\ttag,P,Q)$ with $\typs{\ttag}{P,Q}$.

\begin{definition}[Sequential Bisimulation]\label{d:sb}
  A typed process relation $\R$   is a \emph{sequential bisimulation} if whenever
  $(\ttag,P,Q)\in\R$ and $\rconf{\ttag}{P}\strans{\mu}\rconf{\ttag'}{P'}$, 
then  one of the two following clauses holds:
  \begin{enumerate}
\item there is $Q'$ such that $Q \wtrans{\hat\mu}Q'$ and $(\ttag',P',Q')\in\R$; 

\item $\mu = \earlyinpi{a}{\many{b}}$ and there is $Q'$ such that $Q|\outpi{a}{\many{b}} \wtrans{}Q'$
with $(\ttag',P',Q')\in\R$.
\end{enumerate}
Moreover, the  converse of (1) and (2) holds on the transitions from $Q$.
Processes 
 $P$ and $Q$ are \emph{sequentially bisimilar at $\ttag$}, written $P \wbiss{\ttag} Q$, if
  $(\ttag, P, Q)\in \R$ for some sequential bisimulation $\R$.
\end{definition}

In clause  (2), $Q
| \outpi a b$ is well-typed, be $a$ an input- or output-controlled name.
Clauses (1) and (2) are the same as for ordinary bisimilarity $\wAbis$  (Definition~\ref{d:bisa});
typing however prevents certain transitions to be considered as challenge transitions in
the  bisimulation game. Thus the resulting bisimilarity becomes coarser. 


Ordinary  bisimilarity is included in the sequential one 
(the inclusion is strict, see  Section~\ref{ss:exSeq}). 
\iflong
 also a proof technique for barbed equivalence. 
Indeed, in any bisimulation, the subset of well-typed processes
 yields a  sequential  bisimulation.
\fi 
\begin{proposition}
\label{p:asy_seq}
For $\typs{\eta}{P,Q}$,  if  
 $P \wAbis Q $  then also  
 $P \wbiss{\ttag} Q$.
\end{proposition} 

\iflong
  all the examples in the next section fail
for $\wAbis$.
\fi

		


\begin{theorem}[Soundness]
\label{t:sound_seq}
		If $P\wbiss{\ttag} Q$, then $P \beq{\ttag} Q$.
\end{theorem}

As usual, the proof of Theorem~\ref{t:sound_seq} relies on
the preservation of \wbiss\ttag{} under parallel composition, which
requires some care in order to enforce sequentiality. This is ensured
by typability.
%
Theorem~\ref{t:sound_seq} allows us to use the labelled bisimilarity \wbiss{\ttag} as a proof  technique for typed barbed equivalence.  

This 
proof technique is also complete,
assuming only output-controlled names 
(i.e., the thread may only be exercised  by output particles, not by the input ones). 



\begin{theorem}[Completeness on output-controlled names]\label{t:completeness}
		For all image-finite processes $P,Q$ that only use
    output-controlled names, and for all \ttag, if $P \beq{\ttag} Q$ then $P \wbiss{\ttag} Q$.
\end{theorem}

The completeness proof can be found  in~\cite{HPS:lics21:long}. 
While the overall structure of the proof is standard, 
the technical details  are specific to sequentiality. 
As usual, we rely on a
stratification of bisimilarity and approximants
 $\wbissn{\ttag}{n}$, 
\iflong 

 for the $n$th  of
\wbiss\ttag{} (the definition is standard, and is given in Appendix). As usual, \wbiss\ttag{} coincides with the intersection of its approximants). 
\fi
%
%
%
and reason by contradiction to show that if
$\typs{\ttag}{P,Q}$ and $P \not\wbissn{\ttag}n Q$,
then there is a $\conTT{\ttag'}{\ttag}$ static context $E$ such that $E[P] \not\bbis{\ttag'} E[Q]$.
\iflong
Because of typing, the definition of such $E$ requires some care.
\fi
The case  $\ttag=0$ (tested processes are inactive) is  rather standard: 
the context 
$E$ is of the
form $\respi{\many{x}}(\hole | \out{z} | z.R)$, for some fresh
$z$, and some ``tester process'' $R$. The barb at $z$ allows us  to detect when the tested
process  interacts with $R$.

The delicate case is when  $\ttag=1$ (tested processes are active): the context must be
inactive and hence 
 cannot have an unguarded
output at $z$.
We use in this case a
\iflong  $\conTT{1}{1}$
\fi context of the form
$E \defi \respi{\many{x}}(\hole | G_R+G)$.
%
Process  $G_R$ is the tester process, and $G$ is
 $\sum_{y\in S} \inpi y{\many{y'}}.\out{z}$, defined for some fresh $z$ and some set $S$
 containing $\fnames P\cup\fnames Q$. 
$G$ satisfies the following
 property: for any  $P_0$ and for any $x$,
 if $\typps{1}{P_0\wbarb{\out{x}}}$ then $\typps{1}{P_0 | G
    \wbarb{\out{z}}}$.
  Thus, as soon as $P_0$ exhibits some barb, we have
  $\typps{1}{E[P_0] \wbarb{\out{z}}}$, and $P_0$ cannot interact with $R$
  without removing  the barb at $z$, which allows us to reason as  in the case $\ttag=0$.

The proof schema above does not apply if input-controlled names
are allowed, intuitively because  in this case the processes being tested may 
be active and perform an input (at an input-controlled name), thus maintaining the
thread; both before and after the transition the testing context is passive and hence
unable to signal, with appropriate barbs, which interaction occurred. 

\subsection{Examples}
\label{ss:exSeq}

With respect to ordinary bisimilarity,
in sequential   bisimilarity (\wbiss{\ttag})   fewer challenges are allowed. 
This  may both mean that certain processes, otherwise distinguishable,  become equal, and
that certain equalities are simpler to prove because  
 the state space of the processes to be examined is reduced.
 We 
 present some equalities of the first kind (valid for  \wbiss\ttag{}
only). In
Section~\ref{s:ex:ref}, we also   show a refinement of \wbiss{\ttag} useful for reasoning about references.

\subsubsection{Basic examples}
In the type system, $\nil$ is inactive~---- without the thread. 
We write $\nil_1 $ to abbreviate $ \respi{x}(\out{x})$ (an active process without 
transitions). 

\begin{example}
\label{exa:singu}
While a component of a system is active,  other components  cannot be
observed. Thus, if the active component keeps the thread, 
the existence of other components is irrelevant. 
Indeed we have, for any  $R, Q$ inactive: 
	$$R | \nil_1 \wbiss{1} R | \respi{x}(\out{x} | !x.\out{x}) \wbiss{1}  Q |
        \respi{x}(\out{x} | !x.\out{x}) \wbiss{1}   \nil_1\iflong,\fi$$
\iflong        More generally, say that a process $P$ is \emph{singular} if it never releases the thread;
that is, the set of singular processes is the largest set ${{\cal T}}$ of processes  
such that  for all $P \in {\cal T}$ we have 
$\typs{1}{P}$ and whenever $\sttrans{1}{P}{\mu}{P'}$ then $\mu$ is not an output
and also $P' \in {\cal T}$.
Then for all $P \in {\cal T}$, and for all processes $Q_1,Q_2$ with 
$\typs{0}{Q_1,Q_2}$ we have
\[ 
  Q_1 | P \wbiss{1} Q_2 | P
  .
\] 
The relation 
containing all such pairs 
$ (Q_1 | P, Q_2 | P)$ of processes 
is a sequential  bisimulation.
\fi
	\end{example}

\begin{example}
\label{ex:exp} 
An unguarded occurrence of an input at an input-controlled name becomes the only observation that can be
made in a process. 
This yields the following equalities
		\begin{mathpar}
\begin{array}{rcl}
			u.P | x.Q &\wbiss{1}&  u.(P | x.Q)
\\
			u.P | \out{v} &\wbiss{1}& u.(P | \out{v})  \hskip 1cm \text{ for }u\neq v
\end{array}
 \end{mathpar}
\iflong For the first equality, one shows that 
$$\{ (1,	u.P | x.Q , u.(P | x.Q)) \} \cup \Id \, , $$ 
where $\Id$ is the (typed) identity relation, 
is a    sequential  bisimulation. One  proves the second
equality similarly.
\fi	\end{example}

\iflong 
	By sequentiality, it is also not possible to access parts of
        processes that would require a simultaneous/parallel
        reception. The following example illustrates this.
	\fi
	\begin{example} 
Consider the process 
	$$P \defi \respi{y',z'}(!x.(z'.\out{z}|\out{y'}) | !y.\out{z'}).$$
The output at $z$ becomes observable
if both an input at $x$ and an input at $y$ are consumed, so that the
internal reduction 
at $z'$  can take place. However the input at $x$ acquires the thread, preventing a
further immediate input at $y$; similarly for the input at $y$. 
Indeed we have
$ P  
	\wbiss{0} x.\nil_1 + y.\nil_1 \iflong\defi Q \fi
$.
\iflong 
To prove the equality, we can use  
$ \{P, Q \} \cup    	\wbiss{1}$, which is easily proved to be a sequential
 bisimulation. The derivatives of $P$ and  
$Q$ are singular (and stable, i.e., unable to reduce further) processes;
therefore they are in \wbiss{1}, as discussed in Example~\ref{exa:singu}.
    	
Under the ordinary
 bisimilarity, 
 $P$ and $Q$ are distinguished 
because  the sequence of transitions  $P \strans{x}\strans{y}\strans{\out{z_3}}$ cannot be matched by $Q$.
\fi      \end{example} 

%
%
%
%
%

\iflong 
\begin{example}
This example informally discusses 
why sequentiality can help reducing the  number of pairs of processes
to examine  in a bisimulation proof. 
Suppose we wish to prove 
 the equality between the two  processes 
\[ 
\begin{array}{rcl}
P_1 &\defi &    \outC x |  
\inpC  y. R |  \inpC  u .Q 
\\ 
P_2 &\defi &    \outC x |
 \inpC  u.Q | \inpC  y .R 
\end{array}
 \]
and such processes are typable under the sequentiality system. 
The difference between $P_1$ and $P_2$ comes from a commutativity of parallel composition. 
We may therefore use  ordinary  bisimilarity, as this  is a sound proof technique  for typed barbed
equivalence.
One may prove, more generally,  that parallel composition is commutative and derive the
equality. However suppose we wish to  use the bisimulation method,  concretely, on $P$ and $Q$. 
The two processes have $3$ initial transitions (with labels $\outC x, y$, and $u$), and the subtrees of transitions
emanating from the such derivatives 
have similar size. 
Under ordinary  bisimulation, we have to examine all  the states in the $3$ subtrees. 

Under  sequential  bisimulation, however, only the input at $y$ is observable 
(it is the only one carrying the thread), thus removing $2$ of the $3$ initial subtrees. 
Further pruning may be possible later on, again exploiting the fact that under
sequentiality only certain transitions are observable. 
\end{example} 
\fi
\subsubsection{Examples with references}\label{s:ex:ref}

We now consider a few examples involving  references. 
For this, we use the standard encoding of references into \Api, and we
enhance the 
bisimilarity for sequentiality so to take references into account. 

We use $n,m,...$ to range over the entities stored in references
(which can be names or values belonging to a first-order data type like
booleans and integers) and placeholders for them. Name $\ell$ is used to represent a reference.

In \Api, a 
 reference $\ell$  holding
a value $n$ is
represented as an output particle  \outpi{\ell}n. 
A process that contains a reference $\ell$
should have, at any time,  exactly one unguarded output  at $\ell$, meaning that at any
time the reference has a unique value. We say that in this case $\ell$ \emph{is
accessible}.  
The read and write operations on $\ell$ are
written as follows:
\[
\begin{array}{rcl}
  \rreadL{m}.R & \defi &  \inpi{\ell}{m}.(\outpi\ell m | R) \\
  \rwriteL{n}.R & \defi & \inpi\ell{m'}.(\outpi\ell n | R)~\mbox{ for }m'\notin\fnames R
\end{array} 
\]
Thus a  name $\ell$ used to encode a reference is input-controlled, as an  action on a
reference is represented by an input at $\ell$ |
we use  $\ell$ rather than $u,v,...$ to stress the fact that names
used to represent references obey constraints that go beyond input-control.

Proof techniques for the representation
of references in \Api have been studied  in \cite{DBLP:conf/concur/HirschkoffPS20}. 
Adopting them requires enhancing our type system  with information about 
references, which simply consists in  declaring which names represent references. 
 In the definition of barbed equivalence, the main constraint is that
the tested context should make sure that all existing reference names  are accessible.
To reason about references, several definitions of labelled bisimilarity are presented in
\cite{DBLP:conf/concur/HirschkoffPS20}, varying on the forms of constraints imposed   on transitions. 
Here we  only import the simplest such constraint: it forbids observations of 
input transitions $P \arr{\earlyinpi \ell n}P'$ at a  reference name $\ell$ when
$\ell$ is accessible in 
 $P$ (i.e., an unguarded output at $\ell$ occurs in $P$).   Such a 
constraint represents the fact that an observer may not pretend to own a
 reference when the reference is accessible in the process.

Formally, with  the addition of references,  judgements in the type system become of the form 
$\typsR S \eta P$, where $S$ is a finite set of reference names, meaning that 
$\typs \eta P$ holds 
and that   $S$ is the set of  accessible reference names in $P$. 
The definition of type-allowed transitions, 
$\sttransR S {\ttag}{P}{\mu}{P'}$,
is the same as before (Definition~\ref{d:tat}) with the addition, in clause (3),
 of the
constraint
\begin{center}
$ $ \hfill  if $\mu$ is an input  $ \earlyinpi{\ell}{n}$ at a reference name $\ell$ then $\ell\not\in
 S$.   \hfill $(*) $
\end{center}  
 Finally the definition of \emph{sequential bisimilarity  with references at $(S, \ttag)$}, written 
$\wbRS{S}{\ttag}$ is the same as  that of sequential bisimilarity (Definition~\ref{d:sb}),
just using $\typsR S \ttag {P,Q}$
and 
$\sttransR S {\ttag}{P}{\mu}{P'}$ in place of 
$\typs{\ttag}{P,Q}$
and
 $\sttrans{\ttag}{P}{\mu}{P'}$. 

It is straightforward to extend the soundness proof for 
sequential bisimilarity w.r.t.\ 
barbed equivalence (Theorem~\ref{t:sound_seq}) to the case of 
sequential bisimilarity with references. 
\iflong
We refer to \cite{DBLP:conf/concur/HirschkoffPS20} for further details on
proof techniques for references in  \Api. 
\fi

\begin{example}\label{e:rewr}
This example shows that reading or writing on a global reference is not subject to
interferences from the outside, as these operations require the thread:
$$
\begin{array}{rcl}
\outpi{\ell}{n} | \rreadL{m}.R &\wbRS{\ell}{1}& \outpi{\ell}{n} | R\sub n m\\
\outpi{\ell}{n} | \rwriteL{m}.R &\wbRS{\ell}{1}& \outpi{\ell}{m} | R
\end{array}$$
Indeed, in each law, if $P$ (resp. $Q$) is the process on the left-hand (resp. right-hand) side,
then  the relation $\{((\ell ; 1),P,Q)\}\cup \Id$ is a sequential  bisimulation, when taking
 the constraint $(*)$ for references into account.
\end{example}

\begin{example}[Fetch-and-add, swap]
We consider  fetch-and-add and  swap operations, 
often found in operating systems. 
The first, written
 $\rfaaL{n}{m}$
 atomically increments  by $n$  the content of the reference $\ell$, 
 and returns the original value as $m$; the second, written 
  $\rswapL{n}{m}$, atomically sets the content
  of 
  $\ell$
  to $n$ and returns the original value as $m$:
\[
\begin{array}{rcl}
	\rfaaL{n}{m}.R  & \defi & \inpi{\ell}{m}.(\outpi{\ell}{m+n} | R)
\\
	\rswapL{n}{m}.R & \defi & \inpi{\ell}{m}.(\outpi{\ell}{n} | R)
\end{array}
\]
These operations may be mimicked by a combination of 
read and write operations (we take $m'\notin\fnames R$):
\[\begin{array}{rcl}
	\rfaaD{n}{m}.R  & \defi & 
\rreadL{m}.\rwriteL{m+n}. R \\
& = &
\inpi{\ell}{m}.(\outpi{\ell}{m} |
         \inpi\ell {m'}.(\outpi{\ell}{m+n} | R)) 
	\\
      \rswapD{n}{m}.R & \defi &
\rreadL{m}.\rwriteL{n}. R \\
& = &
\inpi{\ell}{m}.(\outpi{\ell}{m} |
         \inpi\ell {m'}.(\outpi{\ell}{n} | R)) 
\end{array}\]
For this mimicking to be correct, sequentiality is necessary. 
To see this, consider the simple case when $R \defi \outpi c m$. 
In the ordinary \Api,  processes 
$	\rswapL{n}{m}.R$ and 
$	\rswapD{n}{m}.R$ are distinguished, 
intuitively because the observer is capable of  counting the two inputs and the two outputs at
$\ell$ in   $	\rswapD{n}{m}.R$ (against only one in 
$	\rswapL{n}{m}.R$) and/or is capable of detecting the output 
$\outpi{\ell}{m}$ in 
$	\rswapD{n}{m}.R$. 

The processes are also distinguished with the proof techniques for references in~\cite{DBLP:conf/concur/HirschkoffPS20}, intuitively because, after  the initial input 
$\inpi{\ell}{m}$ (whereby the processes read  the content of the reference),
 an observer may interact with the derivative of 
$	\rswapD{n}{m}.R$ and use its output 
$\outpi{\ell}{m}$  
 so to know the value that had been read. Such an observation  is not possible with 
$	\rswapL{n}{m}.R$. 

In contrast, the two processes are equal if we take sequentiality into account. 
That is, we have: 
$$	\rswapL{n}{m}.R 
\wbRS{\emptyset}{1}
	\rswapD{n}{m}.R$$
This is proved by showing that the relation
        $$
\begin{array}{l}
\cup_{m'} \{((\ell;1), \outpi{\ell}{n} | R\sub{m'} m,~ \outpi{\ell}{m'} | \rwriteL{n}.R \sub{m'} m)
\\[3pt]
\cup 
\; \Id\; \cup \; \{((\emptyset;1),\; \:  \rswapL{n}{m}.R, \; \: \rswapD{n}{m}.R)\} 
\end{array}
 $$
%
is a sequential  bisimulation.
The equivalence between
$	\rfaaL{n}{m}.R$ and 
$	\rfaaD{n}{m}.R$ is established using a similar relation.
\end{example} 


\begin{example}[Optimised access]\label{e:optim:access}
	Two consecutive read and/or write operations can be
        transformed into  an equivalent single operation.
\[
\begin{array}{rcl}
		\rwriteL{n}.\rwriteL{m}.R &  \wbRS{\emptyset}{1} &  \rwriteL{m}.R
\\
		\rwriteL{n}.\rreadL{m}.R & \wbRS{\emptyset}{1}&  \rwriteL{n}.R\sub n m
\\
		\rreadL{m}.\rreadL{m'}.R & \wbRS{\emptyset}{1}& \rreadL{m}.R\sub{m}{m'}
\end{array}
\]
For the first equality, one shows that
\begin{align*}
	\Id\;  \cup \;&\{((\emptyset;1), \rwriteL{n}.\rwriteL{m}.R,\; \: \rwriteL{m}.R)\}\\
 \cup \;	&\{((\ell;1), \outpi{\ell}{n} | \rwriteL{m}.R,\; \: \outpi{\ell}{m} | R)\}
\end{align*}
is a sequential  bisimulation.
The second law is treated similarly.  In both cases, the relation exhibited is finite.

For the third equality, one defines $\R$ as
        $$
	\cup_{n}\set{((\ell;1), \outpi{\ell}{n}|\rreadL{m'}.R\sub{n}{m}, \outpi{\ell}{n} | R\sub{n,n}{m,m'}
		)}$$
Then $\{((\emptyset;1), \rreadL{m}.\rreadL{m'}.R,\;
        \rreadL{m}.R\sub{m}{m'})\}$
$
\cup\;\R \; \cup \; \Id$
       is a sequential  bisimulation.
\end{example}






\section{Well-bracketing}\label{s:wb}


\subsection{Type System}

We now go beyond sequentiality, so to handle well-bracketing.  In
 languages without  control operators, this means
that return-call interactions among terms follow a stack-based discipline.
%

Intuitively, a well-bracketed system is a sequential system 
offering services.
When  interrogated, a service, say $A$,  acquires the thread and is supposed to return a final result 
(unless the computation diverges)  thus releasing the thread. 
During its computation, $A$ may however  interrogate another service, say $B$,
which, upon completion of its computation, will return the result to $A$. In a similar
manner,  the service $B$,
 during its computation,  may call yet another service $C$, and will wait for the return
 from $C$ before resuming its computation. $B$ may also delegate to $C$ the task of
 returning a result to $A$. In any case, the `return' obligation may not be thrown away or
 duplicated. 

The implementation of this policy requires 
 \emph{continuation names}.  For instance, when calling $B$,  process $A$   transmits a fresh name, say
$p$, that will be used by $B$ (or other processes delegated by $B$) to return the result
to $A$. Moreover, $A$ waits for such a result, via an input at $p$. Therefore continuation
names are \emph{linear}~\cite{DBLP:journals/toplas/KobayashiPT99}~--- they may only be used once~--- and
\emph{input receptive}  \cite{DBLP:journals/tcs/Sangiorgi99}~---
the  input-end of the name must be made available  as soon as the name is created;
and they are output-controlled: they carry the thread in output.

In short,  the `well-bracketing' type system defined in this section
 refines the type discipline for sequentiality by
adding linear-receptive names and enforcing a stack discipline on the
usage of such names.   
Proof techniques for well-bracketing will be studied in Section~\ref{s:wb:bis}.

\begin{figure*}[h!]
	\begin{mathpar}
          \trans{wb-Out1}\inferrule{ }{\typr{\Tag p \OO}{\outpi{p}{\many a}}}
		\and
		\trans{wb-Out2}\inferrule{ }{\typr{\Tag p \OO}{\outpi{x}{\many a,p}}}
		\and
		\trans{wb-Out3}\inferrule{ }{\typr{\emptyset}{\outpi{u}{\many a}}}
		\\
		\trans{wb-Inp1}\inferrule{\typr{\Tag p\OO}{P}\and p\neq q}
		{\typr{\Tag q\II,\Tag p\OO}{\inpi{q}{\many a}.P}}
		\and
		\trans{wb-Inp2}\inferrule{\typr{\Tag p\OO}{P}}
		{\typr{\emptyset}{\inpi{x}{\many a,p}.P, \; 
				!\inpi{x}{\many	a,p}.P}}
		\and
		\trans{wb-Inp3}\inferrule{\typr{\Tag p\OO}{P}}
		{\typr{\Tag p\OO}{\inpi{u}{\many a}.P}}
		\and
		\trans{wb-Nil}\inferrule{ }{\typr{\emptyset}{\nil}}
		\and
		\trans{wb-Res1}\inferrule{\typr{\xi,\Tag p\OO,\Tag p\II,\sigma'}{P}}
		{\typr{\xi,\sigma'}{\respi{p}P}}
		\and
		\trans{wb-Res2}\inferrule{\typr{\sigma}{P}}
		{\typr{\sigma}{\respi{p}P}}~p\notin\sigma
		\and
		\trans{wb-Res3}\inferrule{\typr{\sigma}{P}}
		{\typr{\sigma}{\respi{a}P}}
		\and
		\trans{wb-Mat}\inferrule{\typr{\emptyset}{P}}
		{\typr{\emptyset}{[a=b]P}}
		\and
		\trans{wb-Par}\inferrule{\typr{\sigma}{P}
			\and
			\typr{\sigma'}{Q}}
                      {\typr{\sigma''}{P|Q}}
                      \inter{\sigma''}{\sigma}{\sigma'}
		\and
		\trans{wb-Tau}\inferrule{\typr{\sigma}{P}}{\typr{\sigma}{\tau.P}}~\sz{\sigma}\leq 1
		\and
		\trans{wb-Sum}\inferrule{\typr{\sigma}{P} \and
			\typr{\sigma}{Q}}{\typr{\sigma}{P + Q}}~\sz{\sigma}\leq 1
	\end{mathpar}
	\caption{Type system for well-bracketing}
	\label{f:typ:rule:int}
\end{figure*}

Thus, with well-bracketing,
we have three kinds of names:  output-controlled names (ranged over by
$x,y,z,...$) and  input-controlled names (ranged over by
$u,v,w...$), as in the previous section; and  continuation names, ranged over by 
$p,q,r...$. As  before, names $a,b,c...$ range over the union of output- an
input-controlled names. 

Continuation names may only be sent at output-controlled names. 
Indeed, any output at an output-controlled name must carry exactly one continuation name. 
Without this constraint the type system for well-bracketing would be
more complex, and 
it is unclear whether it would be useful in practice.
%
By convention, we assume that,
in a tuple of names transmitted over an output-controlled name,  the last name is a continuation name. We
write $\many a,p$ for such a tuple of names.

The type system is presented in Figure~\ref{f:typ:rule:int}.
Judgements are of the form 
\[ \typr \sigma P \]
where $ \sigma $ is a \emph{stack}, namely 
 a sequence of
\emph{input-} and \emph{output-tagged}
 continuation names, in which the input and output
tags alternate, always terminating with an output tag unless the sequence is
 empty: 
\[
  \begin{array}{c}
 \sigma ~ ::= ~   \sigmaO \midd \sigmaI 
    \\
 \sigmaO ~ ::= ~    \Tag p  \OO, \sigmaI 
    \qquad\qquad
 \sigmaI ~ ::= ~    \Tag p  \II, \sigmaO \midd \emptyset 
  \end{array}
\]
Moreover: a  name may appear at most once with a given tag; and, if a name appears
with both tags, then the input occurrence should immediately follow the output occurrence,
as for $p$  in $\Tag{p'}\II, \Tag p \OO, \Tag p \II, \sigma$.
%
We 
write $p\in\sigma$ if name $p$ appears in $\sigma$, and
$\sz{\sigma}$ for the length of the sequence $\sigma$.


Intuitively,  a stack expresses the expected usage of the free continuation names in a
process.  For instance, if 
\[  \typr{\Tag {p_1} \OO, \Tag {p_2} \II, \Tag {p_3} \OO, \Tag {p_3} \II,  \Tag {p_4} \OO} P 
 \] 
then $p_1,..,p_4$ are the free continuation names in $P$; 
among these,  $p_1$ will be used first,  in an output  ($p_1$ may be the subject or an object
of the output); 
then $p_2$ will be used, in an input  interaction with the
environment.  $P$ possesses both
the output and the input capability on $p_3$, and may use both  capabilities by
performing a reduction at $p_3$; or $P$ may transmit the output
capability and then use the input one; the computation for $P$ terminates with an output at $p_4$.
 This behaviour however concerns only the free continuation names of $P$: at any time
 when an output usage is expected, $P$ may decide to create a new continuation name and
 send it out, maintaining its input end. 
The Subject Reduction Theorem~\ref{t:SR:wb} will   formalise 
the  behaviour concerning   continuations names in stacks.


As simple examples of typing, we can derive 
$$\typr{\Tag p\OO}{\outpi pa}
\quad \mbox{and}\quad  \typr{\Tag p\OO}{\inpi ua.\outpi pa}$$
In the latter typing, by rule \trans{wb-Inp3}, an input at an input-controlled name has
the thread, and does not affect the stack because $u$ is not a continuation name.

The same stack can be used to type a
process that invokes a
service at $x$ before sending the result at $p$, as in
$$\typr{\Tag p\OO}{\respi q(\outpi x{b,q} | \inpi qc.\outpi pc)}$$
where $q$ is a  fresh continuation name created when calling $x$. 
To type the process without the 
 restriction at $q$, the stack should mention the input and output
 capabilities for 
 $q$:
$$\typr{\Tag q\OO,\Tag q\II,\Tag p\OO}{\outpi x{b,q} | \inpi qc.\outpi
  pc}$$
For another example, the process $$P_0 \defi \inpi pa.\outpi {p'}a | \inpi qb.\outpi {q'}b$$ can
be typed using two stacks: we have both \typr{\Tag p\II,\Tag {p'}\OO,\Tag
  q\II,\Tag{q'}\OO}{P_0} and 
 \typr{\Tag q\II,\Tag {q'}\OO,\Tag
   p\II,\Tag{p'}\OO}{P_0}. The choice of the stack depends on whether
 the call answered at $p$ has been made before or after the call answered at 
 $q$.

 We comment on  the rules of the type system. 
In \trans{wb-Out1} and \trans{wb-Out2} the obligation in the stack is fulfilled
(the output capability on the only name in the stack is
used in \trans{wb-Out1} and 
 transmitted in \trans{wb-Out2}). 
 \iflong
 Note, in \trans{wb-Out2}, that 
an output at  an output-controlled name  $x$, as  in the system with sequentiality, passes the thread to another
process (i.e., it is like calling a service $x$).
\\COMMENT: not clear that \trans{wb-Out2} actually shows that.\\
\fi
As explained above, the last name in the tuple transmitted at $x$ is
a continuation name, and the only one being transmitted (to enforce
the stack discipline).
%
In contrast, in \trans{wb-Out3} an output at an input-controlled name does not own the thread and
therefore may not carry continuation names. 
In \trans{wb-Inp1} the input-tagged name on top of the stack is used. 
Rule \trans{wb-Inp2} is the complement of  \trans{wb-Out2}. In  \trans{wb-Inp3}, an input at an
input-controlled name maintains the thread.  
In all  rules for input and $\tau$ prefixes, 
 the stack in the premise of the rules 
  may not contain input-tagged continuation names because 
 their input capability must be unguarded  (as they are receptive names).
The same occurs in  rule   \trans{wb-Sum}, following~\cite{DBLP:journals/tcs/Sangiorgi99} where
 choice on  inputs at receptive names is disallowed (though the constraint  could be relaxed).  
Matching is allowed on plain names, but not on continuation names; this
is typical of type systems where the input and output capabilities on 
names are separate \cite{SanWal}; moreover, no continuation name may appear in
the process underneath, to make sure that the obligations on continuation names are not 
eschewed. 
In  \trans{wb-Res1} a continuation name is created, and then its output and input capabilities
are inserted into the stack. 
In the rule, $\xi,\sigma$  is a decomposition of the stack for 
$\res p P$ where $ \sigma $ is a stack beginning with an output tag; 
hence $\xi$ is either empty or it is of the form $\sigma', \Tag p  \II$,  
(i.e. $\xi$ is an initial prefix of the stack, either empty or ending with an input tag).
\iflong
\ds{thanks Enguerrand for your example: very nice.  I'd consider adding it into the
  explanations, in fact. It is commented here below. 
We should
just explain this, and that  rule Res1 goes beyond ordinary stack, putting Enguerrand's example}\fi

Rule  \trans{wb-Res2} is for continuation names that do not appear in the body of the
restriction (this form of  rule is  common   in type systems for linearity, to simplify the
assertion of  Subject Reduction).   
In rule \trans{wb-Par}, the typing stack is
split to type the two process  components $P_1$ and $P_2$; splitting
of the  typing 
is  usual in type systems  with linearity. Here, however, the split must respect  the order
of the names. That is, the stack in the conclusion should be an interleaving of the two stacks
in the premises,  as by the following definition. 
\begin{definition}[Interleaving]
  
We write $\inter{\sigma_1}{\sigma_2}{\sigma_3}$ if  
(i)  $\sigma_1$ is a
stack,  and (ii) $ \sigma_1 $  is an interleaving of $\sigma_2$ and $\sigma_3$ as
by 
the following inductive rules:
\begin{enumerate}
	\item $\inter{\emptyset}{\emptyset}{\emptyset}$
	\item\label{inter:deux} $\inter{\Tag{p}{\OO},\sigma_1}{\sigma_2}
          {\Tag{p}{\OO},\sigma_3}$ if $\inter{\sigma_1}{\sigma_2}{\sigma_3}$
	\item\label{inter:trois} $\inter{\Tag{p}{\OO},\sigma_1}{\Tag{p}{\OO},\sigma_2}{\sigma_3}$ if $\inter{\sigma_1}{\sigma_2}{\sigma_3}$
	\item the same as~\rref{inter:deux} and~\rref{inter:trois}
          with $\Tag{p}{\II}$ instead of $\Tag p\OO$
\end{enumerate}
\end{definition}

If a name
appears both in $\sigma_2$ and in $\sigma_3$ with the same tag, then
$\theinter{\sigma_2}{\sigma_3}$ may not contain any stack.

\iflong\daniel{I commented here the discussion about examples of processes
  that cannot be typed}\fi

Being stack-ordered means that $p.\out{q} | q.\out{p}$ cannot be typed.
Indeed,  the left process would require  $p$  before $q$ in the stack,
whereas the right process needs the opposite.

In rule \trans{Res1}, having the possibility to add names in the middle of the stack
is mandatory to preserve typability after reduction. Consider for instance:

$\respi{q}(\respi{p}(\outpi{b}{p} | p.\out{q}) | q.\out{p'})
   \strans{\respi{p}\outpi{b}{p}} \respi{q}(p.\out{q} | q.\out{p'})$
To type the derivative of the transition above, we have to use rule \trans{Res1}
with $\xi = \Tag{p}{\II}$ and $\sigma' = \Tag{p'}{\OO}$.


Typability in the type system of Figure~\ref{f:typ:rule:int} implies 
typability in the type system for sequentiality. 
Indeed,  when 
$\typr \sigma P$, 
if the first name in $\sigma $ is output-tagged then $P$ is active, otherwise $P$ is
inactive.  We write $\qseqF $ for the function that `forgets' the well-bracketing information in a stack,
therefore $\seqF \sigma  = 1$ if  $\sigma =  \Tag p  \OO, \sigma'$, for some $p$ and $
\sigma'$, and  $\seqF \sigma  =0 $ otherwise. 

\begin{proposition}
\label{p:erase}
If $\typr \sigma P$ then also 
 $\typs{\seqF \sigma} P$.
\end{proposition} 

In Definition~\ref{d:typr},  
we extend type-allowed transitions to processes with continuation
names.  As previously, we must ensure that the process is typed, and
that the transition is allowed by sequentiality (clauses (1) and
(2) below).  Clause (3) says that the first continuation name observed must be on top of
the stack, and that the input or output capability on a continuation name may not be
exercised by the environment when both capabilities are owned by
the process. 
\begin{definition}
\label{d:typr}
We write $\typpr{\sigma}{P\strans{\mu}P'}$ when
\begin{enumerate}
	\item $\typr{\sigma}{P}$
	\item $\sttrans{\seqF{\sigma}}{P}{\mu}{P'}$ and
	\item if $p\in \fnames\mu$ and $p \in \sigma$, then either
          $\sigma = \Tag{p}{\OO},\sigma'$ or $\sigma =
          \Tag{p}{\II},\sigma'$ for some $\sigma'$;
          moreover, 
          if $p\in{\sigma'}$, then $p$ is not the
          subject of $\mu$.
\end{enumerate}
\end{definition}
We exploit type-allowed transitions to define transitions with stacks, which
make  explicit the evolution of the stack.
\begin{definition}
	We note $\rconf{\sigma}{P} \strans{\mu} \rconf{\sigma'}{P'}$
        when
	$\typpr{\sigma}{P\strans{\mu}P'}$ and
	\begin{enumerate}
		\item\label{sr:out:p} if $\mu = \respi{\many{b}}\outpi{p}{\many{a}}$, then $\sigma = 
		\Tag{p}{\OO},\sigma'$
		\item\label{sr:inp:p} if $\mu = \earlyinpi{p}{\many{a}}$, then $\sigma = \Tag{p}{\II},\sigma'$
		\item\label{sr:out:extr} if $\mu = \respi{\many{c},p}\outpi{a}{\many{b},p}$, then
		$\sigma'=\Tag{p}{\II},\sigma$
		\item\label{sr:out:std} if $\mu = \respi{\many{c}}\outpi{a}{\many{b},p}$, then $\sigma = 
		\Tag{p}{\OO},\sigma'$
		\item\label{sr:inp:oc} if $\mu = \earlyinpi{a}{\many{b},p}$, then $\sigma' = \Tag{p}{\OO},\sigma$
		\item\label{sr:tau} if $\mu = \tau$, then for $\sigma = \Tag{p}{\OO},\Tag{p}{\II},\sigma''$ and
		$p \notin \fnames{P'}$, we have $\sigma' = \sigma''$,
		otherwise $\sigma' = \sigma$.
	\end{enumerate}
\end{definition}
In cases~\rref{sr:out:p}, \rref{sr:out:std} (resp.\ \rref{sr:inp:p}), we 
must have $\sigma = \Tag{p}{\OO},\sigma''$ (resp.\ $\Tag{p}{\II},\sigma''$) by definition
of type-allowed transitions.
In clauses~\rref{sr:out:p} and~\rref{sr:inp:p}, the action is an input
or an output at a continuation name that must be on top of $\sigma$,
and is then removed.
In clause~\rref{sr:out:extr}, the action extrudes a continuation name,
and then, following the stack discipline, the process waits for an
answer on that name.
%
In clause~\rref{sr:out:std}, emitting a free continuation name amounts to
passing the output capability on that name to the environment.
Dually, in clause~\rref{sr:inp:oc}, receiving a continuation name imposes to use it in output.
Finally, in clause~\rref{sr:tau}, a $\tau$ transition may come from an interaction at a
continuation name, in which case $\sigma$ is modified. It can also
come from an interaction at a restricted name  
or from  an internal
choice; in such cases, $\sigma$ is unchanged.

\begin{theorem}[Subject Reduction]
	\label{t:SR:wb}
	If $\typr{\sigma}{P}$ and 
	$\rconf{\sigma}{P}\strans{\mu}\rconf{\sigma'}{P'}$ then $\typr{\sigma'}{P'}$.
\end{theorem} 

\iflong\daniel{I commented here the remark about ``trajectory'' of
  continuation names}\fi

If a process owns both the input and the output capability on a continuation name $p$, then
the environment may not use $p$. Semantically, this is the same  as having a restriction
on $p$ in the process.   
It is therefore safe, in the definition of barbed bisimulation and observability, 
 to assume that all such restrictions  are syntactically present, i.e., 
 there is a single occurrence of 
any free continuation name.
We call \emph{clean} such  processes.

\begin{definition}
\label{d:clean}
A stack $ \sigma $ is  \emph{clean}  
if  no name appears in $ \sigma $ both output- and input-tagged. 
A process $P$ is   \emph{clean} if  $\typr \sigma P$ 
for some clean $\sigma$. 
\end{definition}  
 
On clean processes, typing is preserved by reduction. 

\begin{proposition}
\label{p:clean_red}
If     $\typr  \sigma   
{P }$ for $\sigma$ clean,  then $P \longrightarrow P'$ implies 
       $\typr  \sigma   
{P'}$.
\end{proposition}

Defining barbed bisimulation on clean processes, 
we can use the $\qseqF$ function above to recast observability in the well-bracketing 
system from that in the sequentiality  system: thus,
for $\sigma $ clean, we have $\typpr  \sigma   
{P \Dwa_a}$  (resp.\ $\typpr  \sigma {P \Dwa_p}$) if 
$\typps  {\seqF\sigma} {P \Dwa_a}$  (resp.\ $\typps  {\seqF\sigma} {P \Dwa_p}$).

In the definition of barbed equivalence, the contexts testing the processes must be
clean. 
Writing  $\beq \sigma$  for barbed equivalence at $\sigma$, 
we have 
  $P \beq \sigma 
 Q$  if
 $\typr{ \sigma  }{P,Q}$, and for any clean $\sigma'$ and 
 any $\conTT{\sigma'} \sigma $   static context $E$,  
it holds that  $E[P] \bbis{\sigma'}  E[Q]$ (note that $\sigma$ itself need not be clean).


\subsection{Discreet Processes}

In this section we put forward
the subclass of  
 \emph{discreet} processes, in which all continuation names that are exported must be private, and show  
how to  transform any process into a
discreet one. 
Then, on   discreet processes:
\begin{itemize}
\item[(1)]  
we express 
 a behavioural
property  
 that formalises the stack-like discipline on the
usage of continuation names; 
\item[(2)]  
 we 
 develop
 proof techniques,
 in form of labelled bisimilarities, to reason about the behaviour of
 well-typed processes.
\end{itemize}
 (Concerning (2), while the technical details are quite different, 
we  follow the approach of proof techniques for receptive names
in  \cite{DBLP:journals/tcs/Sangiorgi99}, where the techniques are
first defined on processes where only fresh names may be sent.)

\begin{definition}[Discreet processes]
	A process $P$ is \emph{discreet} if any free continuation name
	$p\in\fnames{P}$ may not appear  in the  object of an output, and,
	in any sub-process $\inpi{x}{\many a,q}.Q$, the same holds for
$q$ in $Q$.
	The definition is extended to contexts, yielding
        \emph{discreet contexts}.
\end{definition}


If $E$ and $P$ are discreet, then so is $E[P]$.
We can  transform all  well-typed processes into discreet   processes using
the law in Lemma~\ref{l:gl} below. The law transforms the output of a \emph{global}
continuation name $p$ into the output of a \emph{local} name  $q$. 
In general, all outputs of continuation names in a process $P$ 
  are local, as a  global output
 corresponds to $P$ 
 delegating   a stack-like obligation to
 another process. In other words, in general 
the transformation of a  non-discreet
process into a discreet  one will modify only  a few outputs of the
initial process. 
%
The law in Lemma~\ref{l:gl}
 is valid for barbed congruence, not just barbed
equivalence, and may therefore be applied to any component of a given
process. 
\begin{lemma} 
\label{l:gl}
	$\outpi{x}{\many{a},p} \bcong{\Tag{p}{\OO}} \respi{q}(\outpi{x}{\many{a},q} | \inpi{q}{\many{b}}.\outpi{p}{\many{b}})$.
\end{lemma}
Thus, \iflong TODO FIX In the light of Lemma~\ref{l:gl},  \fi
in  the definition of barbed equivalence (and congruence) 
it is sufficient to consider 
discreet contexts.

A  discreet process may only export private  continuation names. Dually, 
 the process may only receive fresh continuation names from a discreet context.
 We call
\emph{discreet}  the transitions that  satisfy this property.

\begin{definition}[Discreet transitions]
A typed transition 
	$\typpr{\sigma}{P\strans{\mu}P'}$ is \emph{discreet} if 
	any continuation name in the object of $\mu$ is not free in $\sigma$ (and hence
        also in $P$). 
	
\end{definition}

\begin{lemma}
If $P$ 	is discreet, and $\typpr{\sigma}{P\strans{\mu}P'}$ is discreet,
	then $P'$ is discreet.
	If, moreover, 
        $P$ is clean, then so is $P'$.
\end{lemma}


\subsection{The Well-bracketing property on traces}
Following game semantics \cite{DBLP:journals/iandc/HylandO00}, 
we formalise  well-bracketing, that is, the stack-like behaviour of
continuation names for well-typed processes, using traces of actions. 
In this section, all processes are discreet and clean.
A trace for such a process is obtained from a  sequence of  discreet transitions emanating
from the process, with the expected   freshness conditions to avoid ambiguity among names.

\begin{definition}[Trace]\label{d:trace}
A sequence  of actions $\mu_1, \ldots, \mu_n$ is a  \emph{trace  
for a (discreet and clean) process $P_0$ and a stack $\sigma_0$}  if there are 
 $ \sigma_1,\dots,\sigma_n, P_1,\dots,P_n$
such that  for all  $0\leq j < n$ we have 
	$\rconf{\sigma_j}{P_j}\strans{\mu_{j+1}}\rconf{\sigma_{j+1}}{P_{j+1}}$, where 
the transition is discreet, and moreover all continuation names appearing as object
 in $\mu_{j+1}$ are fresh (i.e., the names may not appear in any
 $\mu_i$ for $i\leq j$).
\end{definition}
The notion of discreet transition already imposes that continuation names in object
position do not appear free in the process. The final
condition in Definition~\ref{d:trace} on continuation names ensures us that for actions like
$\respi{p}\outpi{x}{\many{a},p}$, name $p$ is fresh, and that after an action
$\outpi{p}{\many{a}}$ (thus the only allowed interaction at $p$ has been
played), name $p$ cannot be reintroduced, e.g., in an action
$\earlyinpi{x}{\many a,p}$.
We simply say that 
$\mu_1, \ldots, \mu_n$ is a trace, or is a trace for $P$,
when  the stack or the process are clear from the context.
\iflong
\ds{not good $\mu$ for traces, because used usually for
  actions. Anyhow, the lemma has to be reformulated, as also it is not
clear what $\sigma_0 ,  \sigma_n$ are (wrt $\mu_i$). Introduce the
lemma first and explain what it wants to say (eg the stack-like
behaviour); maybe possible to simplify it talking about 
"suffix" of  a stack"? }
\begin{lemma}
	For all traces $\mu_i$, if $\sigma_0 = \sigma_n$, then for some $\sigma_i'$
	(resp. $\xi_i'$), $\sigma_i = \sigma_i', \sigma_0$ (resp. $\xi_i',\sigma_0$) 
	for all $i$.
\end{lemma}
\fi

The well-bracketing property is best described with the notion of questions and answers.
\begin{definition}
For a  trace $\mu_1, \ldots, \mu_n$, we  set $\mu_i \curvearrowright \mu_j$ if $i < j$ and:
	\begin{enumerate}
		\item either $\mu_i = \respi{\many{c},p}\outpi{a}{\many{b},p}$ and $\mu_j =
		\earlyinpi{p}{\many{a'}}$, 
		\item or $\mu_i = \earlyinpi{a}{\many{b},p}$ and $\mu_j = \respi{\many{c}}
		\outpi{p}{\many{a'}}$.
	\end{enumerate}
	Actions $\mu_i$ (with a continuation name in object position) are called \emph{questions}, while
	actions $\mu_j$ (with a continuation name in subject position) are called \emph{answers}.
\end{definition}
A discreet transition is  either an internal transition, or  a question, or an answer.
A question mentioning a continuation name $p$ 
  is matched by an answer at \iflong the same name\fi $p$. When
  questions and answers are seen
as delimiters (`$[_p$',`$]_p$', different for each continuation name), a well-bracketed trace is a substring
of a Dyck word.
\iflong 
It can be expressed as follows.
\fi

\begin{remark}
	For a discreet transition $\rconf{\sigma}{P}\strans{\mu}\rconf{\sigma'}{P'}$, the value
	$\sz{\sigma'}-\sz{\sigma}$ is 1 for a question, 0 for an internal action, and
	$-1$ for an answer.
\end{remark}

\begin{lemma}[Uniqueness]
	Given a trace $\mu_1,\dots,\mu_n$, 
	if $\mu_i \curvearrowright \mu_j$ and $\mu_{i'} \curvearrowright \mu_{j'}$, then
we have  ($i=i'$ iff $j=j'$).
\end{lemma}

\begin{definition}[Well-bracketing]\label{d:wb}
	A trace $\mu_1,\dots,\mu_n$ is
	\emph{well-bracketed} if for all $i < j$, if $\mu_i$ is a question and $\mu_j$
	is an answer with
        $\mu_i\not\curvearrowright \mu_k$
        and
        $\mu_k \not\curvearrowright \mu_j$
        for all $i < k < j$, then $\mu_i \curvearrowright \mu_j$.
\end{definition}

To prove that all traces are well-bracketed,
\iflong(Proposition~\ref{p:wb}), \fi
we need the following property relating questions and answers to  stacks.

\begin{lemma}
Let $\mu_1, \ldots, \mu_n$ be a trace, and $ \sigma_0, \ldots, \sigma_n$ be the corresponding
stacks, as in
Definition~\ref{d:trace}.
Suppose  $\sigma_0 = \sigma_n$, and for all $i$, $\sz{\sigma_i} > \sz{\sigma_0}$. Then
\iflong
$\mu_1$ is a question, $\mu_n$ is an answer, and
\fi
$\mu_1 \curvearrowright \mu_n$.
\end{lemma}

\begin{prop}\label{p:wb}
	Any trace (as by Definition~\ref{d:trace}) is well-bracketed.
\end{prop}


\subsection{Bisimulation and Full Abstraction}\label{s:wb:bis}

As in Section~\ref{s:seq},
a  \emph{wb-typed relation on processes}  is
a set of triplets $(\sigma,P,Q)$ with $\typr{\sigma}{P,Q}$.
\begin{definition}[WB-Bisimulation]
	A wb-typed relation $\R$ on discreet processes is a \emph{wb-bisimulation} if 
	whenever $(\sigma,P,Q)\in \R$ and $\rconf{\sigma}{P}\strans{\mu}\rconf{\sigma'}{P'}$ 
	is discreet, then one of the three following clauses holds:
        \begin{enumerate}
        \item there is $Q'$ with $Q\wtrans{\hat\mu}Q'$
          and $(\sigma', P', Q')\in\R$
        \item $\mu = \earlyinpi{x}{\many a,p}$ and for some fresh $q$,
          there is $Q'$ with 
          $Q | \respi{q}(\outpi{x}{\many a,q} |
          \inpi{q}{\many b}.\outpi{p}{\many b}) \wtrans{} Q'$
          and  $(\sigma',P',Q')\in\R$
        \item $\mu = \earlyinpi{u}{\many{a}}$ and there is $Q'$
          with $Q | \outpi{u}{\many{a}} \wtrans{} Q'$
          and $(\sigma',P',Q')\in\R$,
        \end{enumerate}
        and symmetrically for the transitions from $Q$.

        Processes $P$ and $Q$ are \emph{wb-bisimilar at $\sigma$}, noted $P \biswb^\sigma Q$, if
	$(\sigma,P,Q)\in\R$ for some wb-bisimulation $\R$.
\end{definition}

Compared to Definition~\ref{d:sb}, the clause for input actions
is here split into two clauses.
In clause (2),  we apply Lemma~\ref{l:gl} to obtain a discreet process.

WB-bisimulation is sound with respect to barbed equivalence for all discreet processes.
The main result concerns preservation by parallel composition:
\begin{lemma}[Parallel composition]
  If $P \biswb^\sigma Q$, then for any discreet process $R$ and stacks
  $\sigma',\sigma''$ such that $\typr{\sigma'}{R}$ and
  $\inter{\sigma''}{\sigma}{\sigma'}$, we have $P|R \biswb^{\sigma''} Q|R$.
\end{lemma}
Note that even if $P,R$ are clean,  $P|R$ needs not be so.

\begin{theorem}[Soundness]
	$\biswb^\sigma \protect{\subseteq} \beq{\sigma}$.
\end{theorem}

To prove soundness, we  show that $\biswb^\sigma$ is preserved by all  discreet static
contexts. By Lemma~\ref{l:gl}, we can then replace any non-discreet context with a 
discreet one.
\iflong
 without changing its behaviour.
\fi

We further refine the coinductive technique given by $\biswb^\sigma$ 
by introducing some  up-to techniques, which make it possible to work
with smaller relations.
	We write $P \redd P'$ when the reduction is \emph{deterministic},
	meaning that whenever $P\strans{\mu}P''$, then $\mu = \tau$ and $P' \equiv P''$.
Similarly, we write $P\wredd P'$ if all reduction steps  are deterministic.
Moreover, for a relation $\R$, we write $(\sigma,P,Q)\in\bisup$ when there exists a stack $\sigma'$, a 
	$\conTT{\sigma}{\sigma'}$ context $E$, and processes $P',Q'$ such that $Q \equiv E[Q']$,
	$P \wredd E[P']$ and $(\sigma', P', Q')\in\R$.

\begin{definition}[Up-to static contexts and up-to deterministic reductions]
\label{d:upto}
	A wb-typed relation $\R$ on discreet processes is a \emph{wb-bisimulation up-to 
	static contexts
	and up-to deterministic reductions} if whenever $(\sigma,P,Q)\in \R$,  for any discreet transition
	$\rconf{\sigma}{P}\strans{\mu}\rconf{\sigma'}{P'}$, one of the
        following clauses holds:
	 \begin{enumerate}
		\item  there is $Q'$  with  $Q\wtrans{\hat\mu}Q'$ and $(\sigma',P',Q')\in\bisup$,
		\item  $\mu = \earlyinpi{x}{\many{a},p}$ and for some fresh $q$, there 
		is $Q'$ with
		$Q | \respi{q}(\outpi{x}{\many{a},q} | q.\out{p}) \wtrans{} Q'$
	and $(\sigma',P',Q')\in\bisup$,
    	\item $\mu = \earlyinpi{u}{\many{a}}$ and, there 
	    is $Q'$ with $Q | \outpi{u}{\many{a}} \wtrans{} Q'$
	    and $(\sigma',P',Q')\in\bisup$,
	\end{enumerate}
and symmetrically for  the transitions from  $Q$.
\end{definition}

\iflong
\engue{Davide est ok pour cette remarque dans une version longue:
Note: on ne peut pas utiliser d'up-to deterministic $\tau$ sur des communications
de noms "illégalement" libres (i.e $\out{p} | p.Q \strans{\tau} Q$) vu que ça change
le typage.}
\fi

\begin{lemma}
	If $\R$ is a wb-bisimulation up-to static contexts and up-to
        deterministic reductions,
	then $(\sigma,P,Q)\in\R$ implies $P\biswb^\sigma Q$.
\end{lemma}

\subsubsection{Completeness}\label{s:compl:wb}

As in Section~\ref{s:seq}, we prove completeness for processes that
only use output-controlled names.

\begin{theorem}[Completeness]\label{t:completeness:wb}
	For all image-finite, discreet and clean processes $P,Q$ that only use 
	output-controlled names, and for all $\sigma$, if $P \beq{\sigma} Q$ then $P \biswb^\sigma Q$.
\end{theorem}
As for Theorem~\ref{t:completeness}, the crux of the proof is 
defining the discriminating static contexts. The additional difficulty
is related to receptiveness of continuation names: we cannot use $z.R$ or $G_R + T$, as in
Section~\ref{s:seq}, 
when  the
tester process, $R$ or $G_R$, 
contains an input at a free continuation name.

Suppose $\typr{\sigma}{P}$, for $P$ 
 discreet and clean. We  decompose  $\sigma$  as $\xi, \Tag{p_1}{\OO}, \Tag{q_1}{\II}, \dots, \Tag{p_{n-1}}{\OO}, \Tag{q_{n-1}}{\II},
\Tag{p_n}{\OO}$ for 
$\xi=\emptyset$ or $\xi=\Tag{q}\II$, and 
 then define, for  fresh
 $q_n$ and $\many{x_i}$,
$$E_{\sigma}^{\many{x_i}} \,\defi\, \respi{\many{p_i},\many{q_i}}(\hole | 
\prod_{i \leq n} \inpi{p_i}{\many{y}}.\outpi{x_i}{\many{y},q_i}).$$
\vskip -.2cm 
\noindent 
We have $\typr{\xi, \Tag{q_n}{\OO}}{E_{\sigma}^{\many{x_i}}[P]}$.
Intuitively,  $E_{\sigma}^{\many{x_i}}$ forwards
information from the $p_i$'s (which are in $\sigma$) to the
$x_i$'s. 
Accordingly, the tester process can use names in \many{x_i} (rather than
 in \many{p_i}), and can use them in input.


Let $\ofnames{- }$ denote the set of free output-controlled names. 
We  distinguish two cases.
If $\xi = \emptyset$, then $P$ is active. To follow the reasoning in
the proof of Theorem~\ref{t:completeness}, we work with $E$ of the form 
$$\respi{\many{x}}(E_{\sigma}^{\many{x_i}} | R
+ \prod_{y\in S} 
\inpi{y}{\many{y'},p}.\outpi{z}{p}),$$ for some  set 
\iflong TOFIX
$S$
of names containing $\ofnames P\cup\ofnames Q$.\fi
$S\supseteq\ofnames P\cup\ofnames Q \cup \many{x_i}$
\iflong\engue{alternatively
  $\ofnames{E^{\many{x_i}}_\sigma[P],E^{\many{x_i}}_\sigma[Q]}
  \subseteq S$} \fi
and fresh $z$.

If $\xi = \Tag{q}{\II}$, then $P$ is inactive. 
We reason with $E$ of the form 
$\respi{\many{x},q}(E_{\sigma}^{\many{x_i}} | \outpi{z}{q} | \inpi{z}{q'}.R)$.
By typing, only the continuation name  $q'$ received at $z$ may appear free in
$R$. Such $q'$ will be instantiated with $q$ and then $R$ will use it to test 
the input at $q$ from the tested processes. 
(A restriction on $q$  is needed, as 
 the overall process has to  be clean.)

In both cases, the resulting $E$ is a $\conTT{\Tag{q_n}{\OO}}{\sigma}$ static context where
$q_n$ is a fresh continuation name.
More details on the proof can be found in~\cite{HPS:lics21:long}. 

\subsubsection{An Example}\label{s:awkward}
\label{ss:exWB}

We explain how the techniques we have introduced allow us to reason
about a well-known
example,
the \textit{well-bracketed state change}
(sometimes called `awkward',  or `very awkward', example)~\cite{DBLP:conf/popl/AhmedDR09,DBLP:journals/jfp/DreyerNB12,DBLP:journals/pacmpl/Jaber20}.
It is usually presented  in ML thus:
\vskip .1cm  
$ $ \hskip -.8cm  \begin{tabular}{rcl}
$M_1$ &  $\defi$ &
\smalltexttt{let
$\ell$ = ref 0 in
fun
y ->}\\
&& \smalltexttt{ ($\ell$ := 0;  y()  ; $\ell$ := 1;  y()  ; !$\ell$) }
\\
 $M_2 $ &  $ \defi$ & 
\smalltexttt{fun
y ->  (y()  ;  y()  ; 1)}
\end{tabular}
\vskip .2cm  
\noindent Function $M_2$ makes two calls to an external function \smalltexttt{y}
and returns $1$. The other term, 
$M_1$,  between the two  calls,  modifies a local reference \smalltexttt{$\ell$}, which  is then used
to return the final result. Intuitively, equivalence between the two
functions holds because: (i) the reference \smalltexttt{$\ell$}
in $M_1$ represents
a local state,  not accessible from 
an external function;
(ii)  
 computation
respects well-bracketing (e.g.,  the language does not have  control operators like call/cc).
\iflong
  (with such an operator, it
 would be possible to reinstall the value \texttt{0} at \texttt{$\ell$}, and
 obtain $0$ as final result).
\fi

Below are the  translations of $M_1$ and $M_2$, following a
standard encoding of functions and references in \Api, and using the
notations for references from 
Section~\ref{s:ex:ref}: 
\vskip -.4cm 
\begin{align*}
  \psapp{\enco{M_1}}{p'} &\defi \respi{x,\ell}(\outpi{\ell}{0} | Q)
                           \quad\mbox{with}\\
  Q &\defi      \outpi{p'}{x} | !\inpi{x}{y,p}.\rwriteL{0}.\respi{q}(\outpi{y}{q} |\\
          &\qquad\qquad
            \; q.\rwriteL{1}.\respi{r}(\outpi{y}{r} | r.\rreadL{n}.\outpi{p}{n}))\\
\psapp{\enco{M_2}}{p'}  &\defi \respi{x}(\outpi{p'}{x} | !\inpi{x}{y,p}.\respi{q}(\outpi{y}{q} |\\
&\qquad\qquad\qquad \qquad\qquad q.\respi{r}(\outpi{y}{r} | r.\outpi{p}{1})))
\end{align*}
\vskip -.2cm 
\noindent 
$\psapp{\enco{M_1}}{p'}$ has a unique  transition, $\psapp{\enco{M_1}}{p'}\strans{\respi{x}\outpi{p'}{x}}
P_1$. Similarly, let $P_2$ be the unique derivative  from $\psapp{\enco{M_2}}{p'}$.
The   equivalence between $\psapp{\enco{M_1}}{p'}$ and
$\psapp{\enco{M_2}}{p'}$ follows immediately from
$P_1\biswb^{\emptyset}P_2$.
To prove the latter,  
%
we  exhibit a relation $\R$  containing the 
triple $(\emptyset, P_1, P_2)$
and  show that $\R$  
is 
a wb-bisimulation  
up-to deterministic reductions and static context. 
\iflong
\ds{Intuitively, $\R$ contains the triples such that ...\\
see if there is a simple explanation above, following how Seguei
does. We can also do without}
\fi
To see 
the importance of well-bracketing, $\R$
contains the triple
$$\begin{array}{l}
	\big(\,\Tag{q_2}{\II}, \Tag{p_2}{\OO}, \Tag{r_1}{\II},
    \Tag{p_1}{\OO},\\
    ~\respi{\ell}(\outpi{\ell}{0} | Q | q_2.\rwriteL{1}.\respi{r_2}(\outpi{y}{r_2}\\
    \qquad\qquad
    | r_2.\rreadL{n}.\outpi{p_2}{n})) | r_1.\rreadL{n}.\outpi{p_1}{n}),\\[.1em]
    ~P_2 | q_2.\respi{r_2}(\outpi{y}{r_2}  | r_2.\outpi{p_2}{1}) | r_1.\outpi{p_1}{1} 
    \,\big)
\end{array}$$
%
\noindent Without the well-bracketing constraint, the first process  in the triple could
perform an input at $r_1$,  an internal transition, and finally  an output
{\outpi{p_1}0}. The second process cannot emit $0$, which would allow us  
to distinguish $P_1$ and $P_2$. 
With well-bracketing, since $r_1$ is not on top of
the stack in the triple, the initial transition on $r_1$ is ruled out.


The details of the definition of $\R$ can be found
in~\cite{HPS:lics21:long}. 
In the same Appendix, we also 
 discuss a simplified example, which exposes the main 
difficulties. The primary simplification consists in
using linear functions\iflong, so that the \Api terms do not need replication and can be finite\fi. 
Some twisting in the ML terms    is 
necessary, as $M_1$ and $M_2$ 
become equivalent~--- even dropping well-bracketing~---  if they can
be used at most once.


\vskip -.3cm 
\section{Related work and conclusions}
\label{s:ccl}
\label{ss:rela}

\vskip -.2cm 

Sequentiality is a form of linearity, hence 
\iflong
the rules for 
\fi
our type system has similarity with, and borrow ideas from,  
\iflong type
\fi
 systems with linear types, in languages  for concurrency or
functional languages, including types for managing locks as in~\cite{DBLP:journals/iandc/Kobayashi02}.
The type system in~\cite{DBLP:conf/popl/LeviS00} ensures one that terms of the
Ambient calculus are single-threaded, a notion similar
 to 
the  sequentiality for \Api examined in this paper. 
The type system in~\cite{DBLP:conf/tlca/BergerHY01} has been designed so to make
  the encoding of PCF into the $\pi$-calculus  fully abstract. 
The system therefore goes beyond sequentiality as described in our
paper. For
instance, the system presents  a form of duality  on types and
 ensures that computations are stateless, hence also
deterministic. Indeed, the only behaviours inhabited by the types are
those in the image of the PCF terms.
Types ensure the uniqueness of the computation thread, and such a thread
is carried by outputs (the thread cannot be carried
by input
processes, as in our system). 
The system  \cite{DBLP:conf/tlca/BergerHY01} has been further refined in~\cite{DBLP:journals/iandc/YoshidaBH04}, adding causality information and
acyclicity
constraints, so to ensure strong normalisation of
well-typed processes. 
The issue  of finding labelled bisimilarity  characterisations of
barbed equivalence or
reduction-closed barbed equivalence is  extensively discussed in \cite{SanWal}; see also 
\cite{DBLP:journals/mscs/HennessyR04} for an  example involving types. 
\iflong
SEE IF WE KEEP THE FOLLOWING SENTENCE
We are not aware of bisimulation-based proof techniques for more
advanced  type
systems such as those in \cite{DBLP:conf/popl/LeviS00,DBLP:conf/tlca/BergerHY01}
mentioned above. 
\fi




Type systems for linearity and receptiveness in the $\pi$-calculus have been introduced in 
\cite{DBLP:journals/toplas/KobayashiPT99,DBLP:journals/iandc/IgarashiK00,DBLP:journals/tcs/Sangiorgi99}.
The way we formulate well-bracketing (Definition~\ref{d:wb}) is inspired by 
 `well-bracketed strategies' in game
semantics~\cite{DBLP:journals/iandc/HylandO00,DBLP:conf/lics/Laird97},
used in functional programming languages and
extensions thereof (they have in turn inspired type systems for $\pi$-calculi with
stack-like information and input/output alternation, e.g.,  \cite{DBLP:conf/tlca/BergerHY01,DBLP:journals/entcs/Honda02}).
The notion of \emph{well-bracketed control flow} is studied in the
field of secure compilation, for a wider class of languages. In works like~\cite{DBLP:journals/pacmpl/SkorstengaardDB19,DBLP:conf/csfw/PatrignaniDP16}, the technique of
fully abstract compilation  guarantees control flow
correctness (and, in particular, well-bracketing)
against low-level attacks.


Several methods have been proposed to establish contextual equivalence
of sequential programs that include higher-order and stateful
computation, including the  above-mentioned game semantics, 
(step-indexed Kripke) logical relations
\cite{DBLP:conf/popl/AhmedDR09,DBLP:journals/jfp/DreyerNB12}, dedicated forms of
bisimulations designed on top of an operational semantics of the
languages~\cite{DBLP:conf/popl/KoutavasW06,DBLP:conf/lics/SangiorgiKS07,DBLP:journals/entcs/KoutavasLS11,DBLP:conf/concur/MadiotPS14,DBLP:conf/fscd/BiernackiLP20}.  
Works
like~\cite{DBLP:journals/pacmpl/Jaber20} or 
algorithmic game semantics~\cite{DBLP:journals/fmsd/MurawskiT18},
aim at automatically establishing contextual equivalences, by
relying on model-checking techniques.





The main goal of this paper was to tailor some of the most prominent proof techniques in
the $\pi$-calculus~--- those based on labelled bisimilarity~---  to the
sequentiality and well-bracketing  disciplines. This is instrumental to the use of the $\pi$-calculus
as a model  of programming languages, as 
sequentiality and well-bracketing are  often found in programming languages or 
subsets of them.   
We have shown the usefulness of our techniques on a number of examples, that have mainly to do
with the representation of functions and store~--- none of the equalities in the examples
is valid  in the ordinary bisimilarity of the calculus.

In Section~\ref{ss:exSeq} we have combined our proof technique for sequentiality with
techniques concerning the representation of references in $\pi$-calculus from
\cite{DBLP:conf/concur/HirschkoffPS20}.  The resulting technique allows us in some cases 
to  reason about
programs with store without an explicit representation of the store (as usually required
in the techniques in the literature,  recalled above).  This avoids 
universal quantifications on the possible values contained in the store, thus  reducing the
size of the relation to consider,  sometimes making them finite.  Further possibilities of
reducing the size of  relations may be possible by defining `up-to
techniques' for our bisimilarities, 
as exemplified by the
up-to technique considered in Definition~\ref{d:upto} and applied in  Section~\ref{ss:exWB}. 

Our treatment of sequentiality raises a few technical  questions that deserve
 further investigation.
We would like to see whether  our proof of completeness
(Theorem~\ref{t:completeness}) could be extended to handle
input-controlled names. Similarly, we do not know whether the result still holds if 
internal choice is  disallowed  in inactive processes.
The usual encoding of an internal choice  $\tau.P + \tau.Q$ in terms of parallel
composition as 
 $\respi
  c\,(\out c|c{}.P|c{}.Q)$, for some fresh $c$, is not applicable because 
  the latter process is active (for instance, the encoding is not valid within  a context
  testing active processes).
Indeed, if the result still holds,  the current
\iflong
  completeness
\fi  proof 
\iflong
 would probably
\else
might
\fi
 require some significant  modifications. 
For similar reasons, it is unclear if and  how our completeness proof could be tuned to handle
reduction-closed variants of barbed equivalence~\cite{HoYo95,SanWal}.



In  the \emph{asynchronous}  $\pi$-calculus considered in this paper, 
 an interaction involves only one prefixed process (the input).  Therefore, in the type
 systems,  this process always
 acquires the control on the thread after the interaction. 
In a 
\emph{synchronous} setting, in contrast,  an interaction involves also an output prefix. 
Hence the type systems could be richer, specifying, for each name, 
where
the control on the thread goes after an interaction at that name. 
The representation of references in Section~\ref{ss:exSeq}, however,  might have to be
revisited as it relies on the asynchronous model.

We have studied proof techniques  for sequentiality and well-bracketing 
in the $\pi$-calculus   based
on labelled bisimilarities. We would like to examine also  the impact of the disciplines on  
algebraic theory and modal logics.

\section*{Acknowledgments}
Prebet acknowledges support from the Universit{\'e} Franco-Italienne,
programme Vinci 2020. 
Sangiorgi  acknowledges support from the
MIUR-PRIN project `Analysis of
Program Analyses' (ASPRA, ID: \texttt{201784YSZ5\_004}).
%


\bibliographystyle{IEEEtran}
\bibliography{bib-func}

\begin{thebibliography}{10}
\providecommand{\url}[1]{#1}
\csname url@samestyle\endcsname
\providecommand{\newblock}{\relax}
\providecommand{\bibinfo}[2]{#2}
\providecommand{\BIBentrySTDinterwordspacing}{\spaceskip=0pt\relax}
\providecommand{\BIBentryALTinterwordstretchfactor}{4}
\providecommand{\BIBentryALTinterwordspacing}{\spaceskip=\fontdimen2\font plus
\BIBentryALTinterwordstretchfactor\fontdimen3\font minus
  \fontdimen4\font\relax}
\providecommand{\BIBforeignlanguage}[2]{{%
\expandafter\ifx\csname l@#1\endcsname\relax
\typeout{** WARNING: IEEEtran.bst: No hyphenation pattern has been}%
\typeout{** loaded for the language `#1'. Using the pattern for}%
\typeout{** the default language instead.}%
\else
\language=\csname l@#1\endcsname
\fi
#2}}
\providecommand{\BIBdecl}{\relax}
\BIBdecl

\bibitem{SanWal}
D.~Sangiorgi and D.~Walker, \emph{The Pi-Calculus - a theory of mobile
  processes}.\hskip 1em plus 0.5em minus 0.4em\relax Cambridge University
  Press, 2001.

\bibitem{DBLP:journals/mscs/PierceS96}
B.~C. Pierce and D.~Sangiorgi, ``Typing and subtyping for mobile processes,''
  \emph{Math. Struct. Comput. Sci.}, vol.~6, no.~5, pp. 409--453, 1996.

\bibitem{DBLP:journals/toplas/KobayashiPT99}
\BIBentryALTinterwordspacing
N.~Kobayashi, B.~C. Pierce, and D.~N. Turner, ``Linearity and the
  pi-calculus,'' \emph{{ACM} Trans. Program. Lang. Syst.}, vol.~21, no.~5, pp.
  914--947, 1999. [Online]. Available:
  \url{https://doi.org/10.1145/330249.330251}
\BIBentrySTDinterwordspacing

\bibitem{DBLP:conf/esop/HondaVK98}
\BIBentryALTinterwordspacing
K.~Honda, V.~T. Vasconcelos, and M.~Kubo, ``Language primitives and type
  discipline for structured communication-based programming,'' in
  \emph{ESOP'98, 7th European Symposium on Programming, 1998, Proceedings},
  ser. Lecture Notes in Computer Science, C.~Hankin, Ed., vol. 1381.\hskip 1em
  plus 0.5em minus 0.4em\relax Springer, 1998, pp. 122--138. [Online].
  Available: \url{https://doi.org/10.1007/BFb0053567}
\BIBentrySTDinterwordspacing

\bibitem{AnconaBB0CDGGGH16}
D.~Ancona, V.~Bono, M.~Bravetti, J.~Campos, G.~Castagna, P.~Deni{\'{e}}lou,
  S.~J. Gay, N.~Gesbert, E.~Giachino, R.~Hu, E.~B. Johnsen, F.~Martins,
  V.~Mascardi, F.~Montesi, R.~Neykova, N.~Ng, L.~Padovani, V.~T. Vasconcelos,
  and N.~Yoshida, ``Behavioral types in programming languages,'' \emph{Found.
  Trends Program. Lang.}, vol.~3, no. 2-3, pp. 95--230, 2016.

\bibitem{DBLP:journals/toplas/Kobayashi98}
\BIBentryALTinterwordspacing
N.~Kobayashi, ``A partially deadlock-free typed process calculus,'' \emph{{ACM}
  Trans. Program. Lang. Syst.}, vol.~20, no.~2, pp. 436--482, 1998. [Online].
  Available: \url{https://doi.org/10.1145/276393.278524}
\BIBentrySTDinterwordspacing

\bibitem{DBLP:journals/iandc/Kobayashi02}
\BIBentryALTinterwordspacing
------, ``A type system for lock-free processes,'' \emph{Inf. Comput.}, vol.
  177, no.~2, pp. 122--159, 2002. [Online]. Available:
  \url{https://doi.org/10.1006/inco.2002.3171}
\BIBentrySTDinterwordspacing

\bibitem{DBLP:journals/iandc/DengS06}
\BIBentryALTinterwordspacing
Y.~Deng and D.~Sangiorgi, ``Ensuring termination by typability,'' \emph{Inf.
  Comput.}, vol. 204, no.~7, pp. 1045--1082, 2006. [Online]. Available:
  \url{https://doi.org/10.1016/j.ic.2006.03.002}
\BIBentrySTDinterwordspacing

\bibitem{DBLP:journals/iandc/YoshidaBH04}
\BIBentryALTinterwordspacing
N.~Yoshida, M.~Berger, and K.~Honda, ``Strong normalisation in the pi
  -calculus,'' \emph{Inf. Comput.}, vol. 191, no.~2, pp. 145--202, 2004.
  [Online]. Available: \url{https://doi.org/10.1016/j.ic.2003.08.004}
\BIBentrySTDinterwordspacing

\bibitem{DBLP:journals/toplas/KobayashiS10}
\BIBentryALTinterwordspacing
N.~Kobayashi and D.~Sangiorgi, ``A hybrid type system for lock-freedom of
  mobile processes,'' \emph{{ACM} Trans. Program. Lang. Syst.}, vol.~32, no.~5,
  pp. 16:1--16:49, 2010. [Online]. Available:
  \url{https://doi.org/10.1145/1745312.1745313}
\BIBentrySTDinterwordspacing

\bibitem{HoYo95}
K.~Honda and N.~Yoshida, ``On reduction-based process semantics,'' \emph{TCS},
  vol. 152, no.~2, pp. 437--486, 1995.

\bibitem{SaWa01}
D.~Sangiorgi and D.~Walker, ``Some results on barbed equivalences in
  pi-calculus,'' in \emph{Proc.\ {CONCUR} '01}, ser. Lecture Notes in Computer
  Science, vol. 2154.\hskip 1em plus 0.5em minus 0.4em\relax Springer Verlag,
  2001.

\bibitem{DBLP:journals/mscs/Milner92}
\BIBentryALTinterwordspacing
R.~Milner, ``Functions as processes,'' \emph{Math. Struct. Comput. Sci.},
  vol.~2, no.~2, pp. 119--141, 1992. [Online]. Available:
  \url{https://doi.org/10.1017/S0960129500001407}
\BIBentrySTDinterwordspacing

\bibitem{DBLP:journals/iandc/Sangiorgi94}
\BIBentryALTinterwordspacing
D.~Sangiorgi, ``The lazy lambda calculus in a concurrency scenario,''
  \emph{Inf. Comput.}, vol. 111, no.~1, pp. 120--153, 1994. [Online].
  Available: \url{https://doi.org/10.1006/inco.1994.1042}
\BIBentrySTDinterwordspacing

\bibitem{DBLP:journals/tcs/Sangiorgi99}
------, ``The name discipline of uniform receptiveness,'' \emph{Theor. Comput.
  Sci.}, vol. 221, no. 1-2, pp. 457--493, 1999.

\bibitem{DBLP:conf/concur/HirschkoffPS20}
\BIBentryALTinterwordspacing
D.~Hirschkoff, E.~Prebet, and D.~Sangiorgi, ``On the representation of
  references in the pi-calculus,'' in \emph{31st International Conference on
  Concurrency Theory, {CONCUR} 2020}, ser. LIPIcs, I.~Konnov and
  L.~Kov{\'{a}}cs, Eds., vol. 171.\hskip 1em plus 0.5em minus 0.4em\relax
  Schloss Dagstuhl - Leibniz-Zentrum f{\"{u}}r Informatik, 2020, pp.
  34:1--34:20. [Online]. Available:
  \url{https://doi.org/10.4230/LIPIcs.CONCUR.2020.34}
\BIBentrySTDinterwordspacing

\bibitem{HPS:lics21:long}
\BIBentryALTinterwordspacing
------, ``{On sequentiality and well-bracketing in the $\pi$-calculus},'' Apr.
  2021, working paper or preprint. [Online]. Available:
  \url{https://hal.archives-ouvertes.fr/hal-03203191}
\BIBentrySTDinterwordspacing

\bibitem{DBLP:journals/tcs/AmadioCS98}
\BIBentryALTinterwordspacing
R.~M. Amadio, I.~Castellani, and D.~Sangiorgi, ``On bisimulations for the
  asynchronous pi-calculus,'' \emph{Theor. Comput. Sci.}, vol. 195, no.~2, pp.
  291--324, 1998. [Online]. Available:
  \url{https://doi.org/10.1016/S0304-3975(97)00223-5}
\BIBentrySTDinterwordspacing

\bibitem{Mil91}
R.~Milner, ``The polyadic $\pi$-calculus: a tutorial,'' {LFCS}, Tech. Rep.
  {ECS--LFCS--91--180}, 1991, {A}lso in {\em Logic and Algebra of
  Specification}, ed.\ F.L.\ Bauer, W.\ Brauer and H.\ Schwichtenberg, Springer
  Verlag, 1993.

\bibitem{DBLP:journals/iandc/HylandO00}
\BIBentryALTinterwordspacing
J.~M.~E. Hyland and C.~L. Ong, ``On full abstraction for {PCF:} i, ii, and
  {III},'' \emph{Inf. Comput.}, vol. 163, no.~2, pp. 285--408, 2000. [Online].
  Available: \url{https://doi.org/10.1006/inco.2000.2917}
\BIBentrySTDinterwordspacing

\bibitem{DBLP:conf/popl/AhmedDR09}
\BIBentryALTinterwordspacing
A.~Ahmed, D.~Dreyer, and A.~Rossberg, ``State-dependent representation
  independence,'' in \emph{Proceedings of the 36th {ACM} {SIGPLAN-SIGACT}
  Symposium on Principles of Programming Languages, {POPL} 2009}, Z.~Shao and
  B.~C. Pierce, Eds.\hskip 1em plus 0.5em minus 0.4em\relax {ACM}, 2009, pp.
  340--353. [Online]. Available: \url{https://doi.org/10.1145/1480881.1480925}
\BIBentrySTDinterwordspacing

\bibitem{DBLP:journals/jfp/DreyerNB12}
\BIBentryALTinterwordspacing
D.~Dreyer, G.~Neis, and L.~Birkedal, ``The impact of higher-order state and
  control effects on local relational reasoning,'' \emph{J. Funct. Program.},
  vol.~22, no. 4-5, pp. 477--528, 2012. [Online]. Available:
  \url{https://doi.org/10.1017/S095679681200024X}
\BIBentrySTDinterwordspacing

\bibitem{DBLP:journals/pacmpl/Jaber20}
\BIBentryALTinterwordspacing
G.~Jaber, ``Syteci: automating contextual equivalence for higher-order programs
  with references,'' \emph{Proc. {ACM} Program. Lang.}, vol.~4, no. {POPL}, pp.
  59:1--59:28, 2020. [Online]. Available: \url{https://doi.org/10.1145/3371127}
\BIBentrySTDinterwordspacing

\bibitem{DBLP:conf/popl/LeviS00}
\BIBentryALTinterwordspacing
F.~Levi and D.~Sangiorgi, ``Controlling interference in ambients,'' in
  \emph{{POPL} 2000, Proceedings of the 27th {ACM} {SIGPLAN-SIGACT} Symposium
  on Principles of Programming Languages, 2000}, M.~N. Wegman and T.~W. Reps,
  Eds.\hskip 1em plus 0.5em minus 0.4em\relax {ACM}, 2000, pp. 352--364.
  [Online]. Available: \url{https://doi.org/10.1145/325694.325741}
\BIBentrySTDinterwordspacing

\bibitem{DBLP:conf/tlca/BergerHY01}
\BIBentryALTinterwordspacing
M.~Berger, K.~Honda, and N.~Yoshida, ``Sequentiality and the pi-calculus,'' in
  \emph{Typed Lambda Calculi and Applications, 5th International Conference,
  {TLCA} 2001, Proceedings}, ser. Lecture Notes in Computer Science,
  S.~Abramsky, Ed., vol. 2044.\hskip 1em plus 0.5em minus 0.4em\relax Springer,
  2001, pp. 29--45. [Online]. Available:
  \url{https://doi.org/10.1007/3-540-45413-6\_7}
\BIBentrySTDinterwordspacing

\bibitem{DBLP:journals/mscs/HennessyR04}
\BIBentryALTinterwordspacing
M.~Hennessy and J.~Rathke, ``Typed behavioural equivalences for processes in
  the presence of subtyping,'' \emph{Math. Struct. Comput. Sci.}, vol.~14,
  no.~5, pp. 651--684, 2004. [Online]. Available:
  \url{https://doi.org/10.1017/S0960129504004281}
\BIBentrySTDinterwordspacing

\bibitem{DBLP:journals/iandc/IgarashiK00}
\BIBentryALTinterwordspacing
A.~Igarashi and N.~Kobayashi, ``Type reconstruction for linear -calculus with
  {I/O} subtyping,'' \emph{Inf. Comput.}, vol. 161, no.~1, pp. 1--44, 2000.
  [Online]. Available: \url{https://doi.org/10.1006/inco.2000.2872}
\BIBentrySTDinterwordspacing

\bibitem{DBLP:conf/lics/Laird97}
\BIBentryALTinterwordspacing
J.~Laird, ``Full abstraction for functional languages with control,'' in
  \emph{Proceedings, 12th Annual {IEEE} Symposium on Logic in Computer Science,
  1997}.\hskip 1em plus 0.5em minus 0.4em\relax {IEEE} Computer Society, 1997,
  pp. 58--67. [Online]. Available:
  \url{https://doi.org/10.1109/LICS.1997.614931}
\BIBentrySTDinterwordspacing

\bibitem{DBLP:journals/entcs/Honda02}
\BIBentryALTinterwordspacing
K.~Honda, ``Processes and games,'' \emph{Electron. Notes Theor. Comput. Sci.},
  vol.~71, pp. 40--69, 2002. [Online]. Available:
  \url{https://doi.org/10.1016/S1571-0661(05)82528-9}
\BIBentrySTDinterwordspacing

\bibitem{DBLP:journals/pacmpl/SkorstengaardDB19}
\BIBentryALTinterwordspacing
L.~Skorstengaard, D.~Devriese, and L.~Birkedal, ``Stktokens: enforcing
  well-bracketed control flow and stack encapsulation using linear
  capabilities,'' \emph{Proc. {ACM} Program. Lang.}, vol.~3, no. {POPL}, pp.
  19:1--19:28, 2019. [Online]. Available: \url{https://doi.org/10.1145/3290332}
\BIBentrySTDinterwordspacing

\bibitem{DBLP:conf/csfw/PatrignaniDP16}
\BIBentryALTinterwordspacing
M.~Patrignani, D.~Devriese, and F.~Piessens, ``On modular and fully-abstract
  compilation,'' in \emph{{IEEE} 29th Computer Security Foundations Symposium,
  {CSF} 2016}.\hskip 1em plus 0.5em minus 0.4em\relax {IEEE} Computer Society,
  2016, pp. 17--30. [Online]. Available:
  \url{https://doi.org/10.1109/CSF.2016.9}
\BIBentrySTDinterwordspacing

\bibitem{DBLP:conf/popl/KoutavasW06}
\BIBentryALTinterwordspacing
V.~Koutavas and M.~Wand, ``Small bisimulations for reasoning about higher-order
  imperative programs,'' in \emph{Proceedings of the 33rd {ACM}
  {SIGPLAN-SIGACT} Symposium on Principles of Programming Languages, {POPL}
  2006}, J.~G. Morrisett and S.~L.~P. Jones, Eds.\hskip 1em plus 0.5em minus
  0.4em\relax {ACM}, 2006, pp. 141--152. [Online]. Available:
  \url{https://doi.org/10.1145/1111037.1111050}
\BIBentrySTDinterwordspacing

\bibitem{DBLP:conf/lics/SangiorgiKS07}
\BIBentryALTinterwordspacing
D.~Sangiorgi, N.~Kobayashi, and E.~Sumii, ``Environmental bisimulations for
  higher-order languages,'' in \emph{22nd {IEEE} Symposium on Logic in Computer
  Science {(LICS} 2007), Proceedings}.\hskip 1em plus 0.5em minus 0.4em\relax
  {IEEE} Computer Society, 2007, pp. 293--302. [Online]. Available:
  \url{https://doi.org/10.1109/LICS.2007.17}
\BIBentrySTDinterwordspacing

\bibitem{DBLP:journals/entcs/KoutavasLS11}
\BIBentryALTinterwordspacing
V.~Koutavas, P.~B. Levy, and E.~Sumii, ``From applicative to environmental
  bisimulation,'' in \emph{Twenty-seventh Conference on the Mathematical
  Foundations of Programming Semantics, {MFPS} 2011}, ser. Electronic Notes in
  Theoretical Computer Science, M.~W. Mislove and J.~Ouaknine, Eds., vol.
  276.\hskip 1em plus 0.5em minus 0.4em\relax Elsevier, 2011, pp. 215--235.
  [Online]. Available: \url{https://doi.org/10.1016/j.entcs.2011.09.023}
\BIBentrySTDinterwordspacing

\bibitem{DBLP:conf/concur/MadiotPS14}
\BIBentryALTinterwordspacing
J.~Madiot, D.~Pous, and D.~Sangiorgi, ``Bisimulations up-to: Beyond first-order
  transition systems,'' in \emph{{CONCUR} 2014 - Concurrency Theory - 25th
  International Conference, {CONCUR} 2014. Proceedings}, ser. Lecture Notes in
  Computer Science, P.~Baldan and D.~Gorla, Eds., vol. 8704.\hskip 1em plus
  0.5em minus 0.4em\relax Springer, 2014, pp. 93--108. [Online]. Available:
  \url{https://doi.org/10.1007/978-3-662-44584-6\_8}
\BIBentrySTDinterwordspacing

\bibitem{DBLP:conf/fscd/BiernackiLP20}
\BIBentryALTinterwordspacing
D.~Biernacki, S.~Lenglet, and P.~Polesiuk, ``A complete normal-form
  bisimilarity for algebraic effects and handlers,'' in \emph{5th International
  Conference on Formal Structures for Computation and Deduction, {FSCD} 2020},
  ser. LIPIcs, Z.~M. Ariola, Ed., vol. 167.\hskip 1em plus 0.5em minus
  0.4em\relax Schloss Dagstuhl - Leibniz-Zentrum f{\"{u}}r Informatik, 2020,
  pp. 7:1--7:22. [Online]. Available:
  \url{https://doi.org/10.4230/LIPIcs.FSCD.2020.7}
\BIBentrySTDinterwordspacing

\bibitem{DBLP:journals/fmsd/MurawskiT18}
\BIBentryALTinterwordspacing
A.~S. Murawski and N.~Tzevelekos, ``Algorithmic games for full ground
  references,'' \emph{Formal Methods Syst. Des.}, vol.~52, no.~3, pp. 277--314,
  2018. [Online]. Available: \url{https://doi.org/10.1007/s10703-017-0292-9}
\BIBentrySTDinterwordspacing

\end{thebibliography}

\end{document}

\appendices



\section{The Asynchronous $\pi$-calculus: Operational Semantics}\label{a:def:api}

We give the definition of structural congruence in
Figure~\ref{f:str:cong}.

\begin{figure}[!h]
	\begin{mathpar}
		P|\nil\equiv P
		\and
		P|Q\equiv Q|P
		\and
		P|(Q|R)\equiv (P|Q)|R
		\and
		G+\nil\equiv G
		\and
		G+G'\equiv G'+G
		\and
		G+(G'+G'')\equiv (G+G')+G''
		\and
		!\inpi{a}{\many b}.P\equiv \inpi{a}{\many b}.P|!\inpi{a}{\many b}.P
		\and
		P|\respi{a}Q \equiv \respi{a}{P|Q} \text{ if }a\notin\fnames{P}
		\and
		\respi{a}\respi{b}{P}\equiv \respi{b}\respi{a}P
		\and
		\respi{a}\nil\equiv\nil
		\and
		[a=a]G \equiv G
	\end{mathpar}
	\caption{Structural congruence in \Api}
	\label{f:str:cong}
\end{figure}

The (early) Labelled Transition System for \Api{} is presented in
Figure~\ref{fig:sos:pil:procs}, actions being defined by the following grammar:
$$\mu ::= \tau \OR \earlyinpi{a}{\many{b}} \OR \respi{\many{c}}\outpi{a}{\many{b}}$$

We have $\fnames{\earlyinpi{a}{\many{b}}}=\set a\cup\many b$, and
$\fnames{\respi{\many{c}}\outpi{a}{\many{b}}}=(\set a\cup\many
b)\setminus\many c$.

\begin{figure*}[!th]
	\begin{mathpar}
		\inferrule*[left=\trans{AInp}]{ }{\inpi{a}{\many{b}}.P \strans{\earlyinpi{a}{\many{c}}}P\{\many{c}/\many{b}\}}
		\and\inferrule*[left=\trans{AOut}]{ }{\outpi{a}{\many{b}} \strans{\outpi{a}{\many{b}}}\nil}
		\and\inferrule*[left=\trans{AOpen}]{P \strans{\respi{\many{c}}\outpi{a}{\many{b}}} P'}{\respi{d}P \strans{\respi{d,\many{c}}\outpi{a}{\many{b}}}P'} \text{if } d\in
		\fnames{\respi{\many{c}}\outpi{a}{\many{b}}}\setminus\set{a}
		\and\inferrule*[left=\trans{ARep}]{P|!P \strans{\mu} P'}{!P \strans{\mu}P'}
		\and\inferrule*[left=\trans{ARes}]{P \strans{\mu}P'}{\respi{a}P
			\strans{\mu}\respi{a}P'}\text{if }a\notin\fnames{\mu}\cup\bnames{\mu}
		\and\inferrule*[left=\trans{APar}]{P \strans{\mu} P'}{P|Q \strans{\mu}P'|Q}\text{if }\bnames\mu \cap \fnames{Q}=\emptyset
		\and\inferrule*[left=\trans{AComm}]{P \strans{\earlyinpi{a}{\many{b}}}P' \and Q \strans{\respi{\many{c}}\outpi{a}{\many{b}}} Q'}{P|Q\strans{\tau}\respi{\many{c}}(P'|Q')}\text{if }\many{c}\cap \fnames{P} = \emptyset
		\and\inferrule*[left=\trans{AMatch}]{P \strans{\mu} P'}{[a=a]P \strans{\mu} P'}
		\and\inferrule*[left=\trans{ASum}]{P \strans{\mu} P'}{P + Q \strans{\mu} P'}
	\end{mathpar}
	\caption{Labelled Transition Semantics for \Api}
	\label{fig:sos:pil:procs}
\end{figure*}

\section{Completeness for sequential bisimilarity}\label{a:compl:seq}

\begin{definition}[Approximants of sequential bisimilarity] We define a sequence $(\wbissn{\ttag}{n})_{n\geq 0}$ of
	typed relations:
	\begin{enumerate}
		\item $P\wbissn{\ttag}{0} Q$ if $\typs{\ttag}{P,Q}$. 
		\item For $1 \leq i$,  relation $\wbissn{\ttag}{i}$ is defined by: 
		$P\wbissn{\ttag}{i} Q$ if whenever $\rconf{\ttag}{P}\strans{\mu}\rconf{\ttag'}{P'}$,
		one of these two clauses holds:
		\begin{description}
			\item[--] there is $Q'$ such that $Q\wtrans{\hat\mu}Q'$ and $P' \wbissn{\ttag'}{i-1} Q'$; 
			\item[--] $\mu = \earlyinpi{a}{b}$ and there is $Q'$ such that $Q|\outpi{a}{b}\wtrans{}Q'$ and $P' \wbissn{\ttag'}{i-1} Q'$,
		\end{description}
		and symmetrically for the transitions of $Q$.
		\item Then $P \wbissn{\ttag}{\omega} Q$ if $P \wbissn{\ttag}{i} Q$ for all $i$.
	\end{enumerate}
\end{definition}
Notice that $\wbissn{\ttag}{0} \supseteq \wbissn{\ttag}{1} \supseteq \dots \supseteq \wbissn{\ttag}{n} \supseteq \dots \supseteq \wbissn{\ttag}{\omega} \supseteq \wbiss{\ttag}$.

\begin{lemma}\label{l:img:eq}
	If $P,Q$ are image-finite, $P\wbissn{\ttag}{\omega} Q$ iff $P \wbiss{\ttag} Q$.
\end{lemma}


We say that a process $P$ is \emph{singular} if it never releases the thread;
that is, the set of singular processes is the largest set ${{\cal T}}$ of processes  
such that  for all $P \in {\cal T}$ we have 
$\typs{1}{P}$ and whenever $\sttrans{1}{P}{\mu}{P'}$ then
$P' \in {\cal T}$.
Singular processes are sequentially equivalent to
$\nil_1$ (an active process without transitions).

When only output-controlled names are used, we can give the following
characterisation of singular processes:
\begin{lemma}\label{l:singular}
  Suppose $P$ is active, and uses only output-controlled names.
  Then $P$ is singular iff there is no $z$ such that $P\Downarrow_{\out{z}}$.

\end{lemma}
This is not true for processes  with input-controlled names: for instance, $u.\out{x}$ is not singular but does not have
a visible barb (it needs to interact on $u$ first).

The following property is the main technical ingredient to derive
completeness.

\begin{prop}\label{p:compl:seq}
  For all $n\geq 0$, if $\typs{\ttag}{P,Q}$, and $P \not\wbissn{\ttag}{n} Q$ and both $P$ and
  $Q$ only use
  output-controlled names and are image-finite, then there exists $R$
  such that for any fresh name $z$, any $\many{x},S$ with
  $\many{x}\subseteq\fnames{P}\cup\fnames{Q}\subseteq S$,
  we define
  $$R_0 \defi\left\{\begin{array}{l l}
          \out{z} | z.R & \mbox{  when }\typs{0}{P,Q}\\
 R + \sum_{y\in S} \inpi{y}{\many{y}'}.\out{z} & \mbox{ when }\typs{1}{P,Q}\end{array}\right.$$
  and we have:
	\begin{enumerate}
		\item either $\respi{\many x}(P' | R_0) \not\bbis{1}
		\respi{\many x}(Q | R_0)$ for all $P'$ such that $P \wtrans{} P'$
		\item or $\respi{\many x}(P | R_0) \not\bbis{1} 
		\respi{\many x}(Q' | R_0)$ for all $Q'$ such that $Q \wtrans{} Q'$.
	\end{enumerate}
\end{prop}
In the second case for the definition of $R_0$, we define $R$ as a guarded process, in
order for the sum to make sense.

\begin{proof}
	We reason by induction on $n$. For $n = 0$, there is nothing to prove.
	
	For $n > 0$, suppose that $P \not\wbissn{\ttag}{n} Q$. Thus, there exists $\mu$
	such that $\rconf{\ttag}{P}\strans{\mu}\rconf{\ttag'}{P'}$ and 
	for all $Q'$ with $Q \wtrans{\hat\mu} Q'$
	(or when $\mu = \earlyinpi{x}{\many{y}}$, all $Q'$ with 
	$Q|\outpi{x}{\many{y}} \wtrans{} Q'$),
	we have $P' \not\wbissn{\ttag'}{n-1} Q'$.
	
	We note $\set{Q_i}$ for $i\in I$ the set of all such
        $Q'$. This set is finite by
	hypothesis. We also write  $S'_I = \bigcup_{i\in I}\fnames{Q_i} \cup
        \fnames{P'}$, which is also a finite set.
	
	We show that clause 1) holds.  For that, we distinguish two
        cases, according to whether $P'$ is singular or not.

        \noindent\textbf{First case: $P'$ is singular}

	Since $P'$ is singular,  $\ttag' = 1$. For all $i$, as
	$P' \not\wbiss{1} Q_i$, we have $Q_i\wbarb{\out{z_i}}$ for
        some $z_i$.
	By definition, $S'_I$ contains all such $z_i$'s.
	
	By Theorem~\ref{l:subred:seq1} (Subject Reduction), there are two possible cases:
	\begin{itemize}
		\item when $\mu = \tau$, we can take $R \defi \nil$.
		We then have to show that for any $z,\many{x},S, Q'$ with $Q \wtrans{} Q'$,
		we have $A \not\bbis{1} B$ where
		\begin{mathpar}
			A\defi \respi{\many x}(P | \nil + \sum_{y\in S} \inpi{y}{\many{y}'}.\out{z})
			\and
			B \defi \respi{\many x}(Q' | \nil + \sum_{y\in S} \inpi{y}{\many{y}'}.\out{z}) 
		\end{mathpar}
		
		We reason by contradiction. We have, since $\mu=\tau$,
                $A \red \respi{\many x}(P' | \sum_{y\in S}
                \inpi{y}{\many{y}'}.\out{z}) \defi A'$ which is also
                singular.
		Thus there is $B'$ such that $B \wred B'$ and \linebreak $A'
   \bbis{1} B'$.

   In such a situation, $B'$ can be of 2 forms:
		\begin{itemize}
			\item $\respi{\many x}(Q_i | \sum_{y\in S} \inpi{y}{\many{y}'}.\out{z})$ for some $i\in I$ (with $Q' \wred Q_i$)
			\item $\respi{\many x}(Q'' | \out{z})$
		\end{itemize}
		The second process clearly is not barbed bisimilar to $A'$.
		For the first one, notice that $\fnames{Q_i} \subseteq \fnames{Q}$ and 
		$\fnames{P'} \subseteq \fnames{P}$. Thus, $S'_I \subseteq S$. As $Q_i
		\wbarb{\out{z_i}}$ and $z_i \in S'_I$;
		this means that $B'\wbarb{\out{z}}$ and thus this process not barbed bisimilar to $A'$.
		
		\item when $\mu = \earlyinpi{x'}{\many{y}'}$, given a fresh name $z'$,  we define $R \defi \outpi{x'}{\many{y}'} | \sum_{y\in S'_I} \inpi{y}{\many{y}'}.\out{z'}$.
		We then have to show that for any $z,\many{x},S, Q'$ with $Q \wtrans{} Q'$
		$$A\defi \respi{\many x}(P | z.R | \out{z}) \not\bbis{1}
		\respi{\many x}(Q' | z.R | \out{z}) \defi B$$
		
		We reason by contradiction. \\
		We have $A \red\red \respi{\many x}(P' | 
		\sum_{y\in S'_I} \inpi{y}{\many{y}'}.\out{z'}) \defi A'$
		which is also singular.
		Thus there should exist $B'$ such that $B \wred B'$ and $A' \bbis{1} B'$.
		We observe that $B'$ can be of 3 forms:
		\begin{itemize}
			\item $\respi{\many x}(Q_i | \sum_{y\in S} \inpi{y}{\many{y}'}.\out{z'})$ for some $i\in I$ (with $Q'|\outpi{x'}{\many{y}'} \wred Q_i$)
			\item $\respi{\many x}(Q'' | z.R | \out{z})$
			\item $\respi{\many x}(Q'' | \out{z'})$
		\end{itemize}
		The latter two are clearly not barbed bisimilar to $A'$.
		For the first case, $Q_i \wbarb{\out{z_i}}$ and $z_i \in S'_I$, 
		this means that $B'\wbarb{\out{z'}}$, and thus $B'$ is not barbed bisimilar to $A'$.
	\end{itemize}

        \noindent\textbf{Second case: $P'$ is not singular.}
	We know in this case that
        either $P'$ is inactive,
	or $P'\wbarb{\out{z}}$ for some $z$ (by Lemma~\ref{l:singular}).

          We note $R_i$ for the process that we obtain, by induction,
          for each pair $(P',Q_i)$.
	\begin{itemize}
		\item when $\mu = \tau$ and $\typs{1}{P'}$, given fresh names $(z_i)_{i\in I}$,  we pose $R \defi \sum_{i\in I} \tau.(R_i + \sum_{y\in S'_I}\inpi{y}{\many{y}'}.\out{z_i})$.
		
		We reason by contradiction. \\
		We have
		$A \red \respi{\many x}(P' | R + \sum_{y\in S} \inpi{y}{\many{y}'}.\out{z}) \defi A'$
		
		Thus there should exist $B'$ such that $B \wred B'$ and $A' \bbis{1} B'$.
		As $P'$ is not singular, we have by
                Lemma~\ref{l:singular} $P'\wbarb{\out{z'}}$ for some
                $z'\in S'_I$ ($P'$ is active because \typs{1}{P'}).
		So $A' \wbarb{\out{z_i}}$ for all $i\in I$
		Therefore $B'$ must be of the form
		$\respi{\many x}(Q_i | R + \sum_{y\in S} \inpi{y}{\many{y}'}.\out{z})$ for some
		$i\in I$ (with $Q' \wred Q_i$)
		
		Now, we have two cases depending on the clause of the
                proposition that holds, by induction, for $(P',Q_i)$:
		\begin{itemize}
			\item if clause 1) holds, then $B' \red B'_i$ with $$B'_i \defi \respi{\many{x}}(Q_i | R_i + \sum_{y\in S'_I}\inpi{y}{\many{y}'}.\out{z_i})$$
			This means there is $A'_i$ with $A' \wred B'_i$ and $A'_i \bbis{1} B'_i$.
			As we do not have $B'_i\wbarb{z_j}$ for $j\neq i$, we must have that
			\begin{itemize}
				\item either $A'_i \equiv \respi{\many{x}}(P'' | 
				R_i + \sum_{y\in S'_I}\inpi{y}{\many{y}'}.\out{z_i})$ with $P' \wred P''$
				\item or $A'_i \equiv \respi{\many{x}}(P'' | 
				\out{z_i})$ for some $P''$
			\end{itemize}
			However the second case is not possible as $B'_i \red \not\wbarb{\out{z_i}}$.
			
			Using structural congruence, we can suppose
                        that in $A'_i$ and $B'_i$, $\many{x}$
			 contains only names from $\fnames{P',Q_i}$.

			We are in a situation where we can apply the induction hypothesis, and 
			this is in contradiction with $A'_i \bbis{1} B'_i$.
                         
			
			\item if clause 2) holds, then $A' \red A'_i$ with 
			$$A'_i \defi \respi{\many{x}}(P' | R_i + \sum_{y\in S'_I}\inpi{y}{\many{y}'}.\out{z_i})$$
			This means there is $B'_i$ with $B' \wred B'_i$ and $A'_i \bbis{1} B'_i$.
			As we do not have $A'_i\wbarb{z_j}$ for $j\neq i$, we must have that
			\begin{itemize}
				\item either $B'_i \equiv \respi{\many{x}}(Q_i' | 
				R_i + \sum_{y\in S'_I}\inpi{y}{\many{y}'}.\out{z_i})$ with $Q_i \wred Q_i'$
				\item or $B'_i \equiv \respi{\many{x}}(Q'' | 
				\out{z_i})$ for some $Q''$
			\end{itemize}
			However the second case is not possible as $A'_i \red \not\wbarb{\out{z_i}}$.
						
			Using structural congruence, we can suppose
                        that in $A'_i$ and $B'_i$, $\many{x}$
			 contains only names from $\fnames{P',Q_i}$.
			
			 We are in a situation where we can apply the induction hypothesis, and 
			 this is in contradiction with $A'_i \bbis{1} B'_i$.
		\end{itemize}
	
		\item when $\mu = \earlyinpi{x'}{\many{y}'}$, given
                  fresh names $(z_i)_{i\in I}$,  we set $R \defi
                  \outpi{x'}{\many{y}'} | \sum_{i\in I} \tau.(R_i +
                  \sum_{y\in S'_I}\inpi{y}{\many{y}'}.\out{z_i})$
                                  and we can conclude.

		\item when $\mu = \tau$ and $\typs{0}{P'}$, given
                  fresh names $(z_i)_{i\in I}$,  we pose $R \defi
                  \sum_{i\in I} \tau.(\out{z_i} | z_i.R_i)$
                  and we can conclude.

		\item when $\mu = \respi{\many{y}^2}\outpi{x'}{\many{y}'}$, given fresh names
		$z',z'', (z_i)_{i\in I}$,
		we note $\many{y}^2 = y^2_1, \dots, y^2_n$ and $\many{y}'=\many{y}^2, y^1_1,
		\dots y^1_m$. 
		
		Then we set
                
                \begin{align*}
                  R \defi&~ \inpi{x'}{\many{x}^2,\many{x}^1}.\big(\,\out{z'}
                  \\&
		| \sum_{j\leq n}\sum_{y''\in\fnames{P,Q}} [x^2_j =
  y'']z'.\out{z''}
                  \\[.1em]&
		+ \sum_{i\in I}[x^1_1 = y^1_1]
		\dots[x^1_m = y^1_m]z'.(\out{z_i} | z_i.R_i)\,\big)
                \end{align*}
                
                and we can conclude.

                
	\end{itemize}
\end{proof}

\begin{proof}[Proof of Theorem~\ref{t:completeness}]
We suppose $P\beq{\ttag}Q$ and $P,Q$ are image-finite.

We reason by contradiction:

If $P\not\wbiss{\ttag} Q$, then by Lemma~\ref{l:img:eq}, $P\not\wbissn{\ttag}{\omega} Q$.

Thus there exists $n$ such that $P\not\wbissn{\ttag}{n} Q$.

By Proposition~\ref{p:compl:seq}, there exists some static context $E$ such that $E[P] \not\bbis{\ttag} E[Q]$.

So $P\not\beq{\ttag}Q$, which is absurd.
\end{proof}

\section{Completeness for WB-bisimulation}\label{a:compl:wb}

The proof of completeness for $\biswb^\sigma$ w.r.t.\ $\beq\sigma$ has
the same overall structure as the proof for sequential
bisimulation. Technically, however, because of the presence of
continuation names, the details are different.

The following lemma is a direct consequence of the usage discipline of
continuation names.

\begin{lemma}\label{l:lin}
	If $\typr{\sigma}{P|Q}$ then
	$$\respi{p}(\outpi{p}{\many{a}} | \inpi{p}{\many{b}}.P | Q) \wbiss{\sigma}
	P\sub{\many{a}}{\many{b}} | Q$$
\end{lemma}
By soundness, this equivalence also holds for barbed equivalence at $\sigma$.

Suppose $\typr{\sigma}{P}$. We recall the definition of
$E^{\many{x_i}}_\sigma$, given in Section~\ref{s:compl:wb}:

With $\sigma=\xi, \Tag{p_1}{\OO}, \Tag{q_1}{\II}, \dots, \Tag{p_{n-1}}{\OO},
\Tag{q_{n-1}}{\II}, \Tag{p_n}{\OO}$ and 
$\xi=\emptyset$ or $\xi=\Tag{q}\II$, we define, for  fresh
$q_n$ and $\many{x_i}$,
$$E_{\sigma}^{\many{x_i}} \,\defi\, \respi{\many{p_i},\many{q_i}}(\hole | 
\prod_{i \leq n} \inpi{p_i}{\many{y}}.\outpi{x_i}{\many{y},q_i}).$$

When $\rconf{\sigma}{P}\strans{\respi{\many{x}}\outpi{p_1}{\many{y}}}\rconf{\sigma'}{P'}$,
then with $\many{x_i} = x_1,\many{x_i}'$
$$\typpr{\Tag{q_n}{\OO}}{E^{\many{x_i}}_\sigma[P] \strans{\tau} 
	\respi{q_1,\many{x}}(E^{\many{x_i}'}_{\sigma'}[P'] | \outpi{x_1}{\many{y},q_1})}$$
By Lemma~\ref{l:lin}, this internal transition does not change the behaviour of the 
process. Thus, in the proof of Proposition~\ref{p:compl:wb}, we reason
modulo this kind of reductions.

As in Appendix~\ref{a:compl:seq}, forbidding input-controlled names
allows us to gain information on the behaviour of singular processes.
\begin{lemma}[Wb-typability and singular processes]
  Suppose a clean active process $P$ does not use input-controlled
  names, and suppose $\typr{\Tag{p}{\OO},\sigma}{P}$.  Then $P$ is not
  singular iff $P\Downarrow{\out{p}}$ or  $P\Downarrow_{\out{z}}$ for some $z$.
	
  Additionally, if $P$ is singular, so is $E[P]$ for any $\rho$ and any
  $\conTT{\rho}{\Tag{p}{\OO},\sigma}$ static context $E$.
\end{lemma}
Under the same hypotheses, this lemma implies that $P$ is not singular
iff $E_\sigma^{\many{x_i}}[P]\wbarb{\out{z}}$ for some $z$.
	
The following proposition is the counterpart, for WB-bisimulation, of
Proposition~\ref{p:compl:seq}. It refers to the stratification of
WB-bisimulation, which is defined like the stratification of
sequential bisimulation.

\begin{prop}\label{p:compl:wb}
	For all $n\geq 0$, $\sigma = \xi, \sigma''$ and $\sz{\xi} \leq 1$, if $\typr{\sigma}{P,Q}$ and $P \not\biswb^{\sigma,n} Q$, then for any fresh names
	$\many{x_i}$ there exists $R$ such that for any fresh $z$ and any $\many{x},S$ 
	with $\many{x}\subseteq \ofnames{P,Q}$ and $\ofnames{P,Q}\cup\many{x_i} \subseteq S$,
	we have:
	\begin{itemize}
		\item When $\xi = \Tag{p}{\II}$ for some $p$:\\
		 We note $R_0 \defi \outpi{z}{p} | \inpi{z}{p}.R$. One of the 
		 following holds:
		\begin{enumerate}
			\item either for all $P'$
			such that $P \wtrans{} P'$
			$$\respi{\many x, p}(E^{\many{x_i}}_{\sigma}[P'] |
			R_0) \not\bbis{\Tag{q_n}{\OO}}
			\respi{\many x, p}(E^{\many{x_i}}_{\sigma}[Q] |
                        R_0)$$
			\item or for all $Q'$ such that $Q \wtrans{} Q'$
			$$\respi{\many x, p}(E^{\many{x_i}}_{\sigma}[P] | R_0) 
			\not\bbis{\Tag{q_n}{\OO}} \respi{\many x, p}(E^{\many{x_i}}_{\sigma}[Q'] | R_0)$$
			
		\end{enumerate}
		\item When $\xi = \emptyset$, we note $R_0\defi R + \sum_{y\in S} 
		\inpi{y}{\many{y}',p}.\outpi{z}{p}$. One of the following holds:
		\begin{enumerate}
			\item either
			$\respi{\many x}(E^{\many{x_i}}_{\sigma}[P'] | R_0) \not\bbis{\Tag{q_n}{\OO}} 
			\respi{\many x}(E^{\many{x_i}}_{\sigma}[Q] | R_0)$ for all $P'$ such that $P \wtrans{} P'$,
			\item or 
			$\respi{\many x}(E^{\many{x_i}}_{\sigma}[P] | R_0) \not\bbis{\Tag{q_n}{\OO}} 
			\respi{\many x}(E^{\many{x_i}}_{\sigma}[Q'] | R_0)$ for all $Q'$ such that $Q \wtrans{} Q'$.
		\end{enumerate}
	\end{itemize}
\end{prop}

The proof of this proposition is structured like the proof of
Proposition~\ref{p:compl:seq}. We therefore do not provide all details, but instead
highlight the technical points that are specific.

\begin{proof}
	We reason by induction on $n$. For $n = 0$, there is nothing to prove.
	
	For $n > 0$, suppose that $P \not\wbissn{\sigma}{n} Q$. Thus, there exists $\mu$
	such that $\rconf{\sigma}{P}\strans{\mu}\rconf{\sigma'}{P'}$ and 
	for all $Q'$ with $Q \wtrans{\hat\mu} Q'$
	(or when $\mu = \earlyinpi{x}{\many{y},p}$, all $Q'$ with 
	\linebreak $Q|\respi{q}(\outpi{x}{\many{y},q} | \inpi{q}{\many{y}'}.\outpi{p}{\many{y}'}) \wtrans{} Q'$ for any fresh $q$),
	we have $P' \not\wbissn{\sigma'}{n-1} Q'$.
	
	We note $\set{Q_i}$ for $i\in I$ the set of all such $Q'$. This set is finite by
	hypothesis. We also write \linebreak $S'_I = \bigcup_{i\in I}\ofnames{Q_i} \cup \ofnames{P'} \cup\many{x_i}$, which is also a finite set.
	
	We show that clause 1) holds.
	
	We distinguish two cases, according to whether $P'$ is singular or not.

	\noindent\textbf{First case: $P'$ is singular.}
	
	Since $P'$ is singular, $\sigma' = \Tag{p_j}{\OO},\sigma_0$
	for some $p_j, \sigma_0$. For all $i$, as
	$P' \not\wbiss{1} Q_i$, we have $E^{\many{x_i}}_{\sigma'}[Q_i]\wbarb{\out{z_i}}$
	for some $z_i \in S'_I$.
		
	By Theorem~\ref{t:SR:wb}, there are two possible cases:
	\begin{itemize}
		\item when $\mu = \tau$ and $\sigma' = \sigma = \Tag{p_1}{\OO},\sigma''$ for
		some $p_1, \sigma''$. We set $R \defi \nil$.	
		We then have to show that for any $z,\many{x},S, Q'$ with $Q \wtrans{} Q'$,
		we have $A \not\bbis{\Tag{q_n}{\OO}} B$ where
		\begin{mathpar}
			A\defi \respi{\many x}(E^{\many{x_i}}_\sigma[P] | \nil + \sum_{y\in S} \inpi{y}{\many{y}',q}.\outpi{z}{q})
			\and
			B \defi \respi{\many x}(E^{\many{x_i}}_\sigma[Q'] | \nil + \sum_{y\in S} \inpi{y}{\many{y}',q}.\outpi{z}{q}) 
		\end{mathpar}
		
		We reason by contradiction.
                We have
		$A \red \respi{\many x}\big(\,E^{\many{x_i}}_\sigma[P'] | \sum_{y\in S} \inpi{y}{\many{y}',q}.\outpi{z}{q}\,\big) \defi A'$ which is also singular.
		Thus there should exist $B'$ such that \linebreak $B
                \wred B'$ and $A' \bbis{\Tag{q_n}{\OO}} B'$.
                
		We observe then that $B'$ can be of 3 forms:
		\begin{itemize}
			\item $\respi{\many x}(E^{\many{x_i}}_\sigma[Q_i] | \sum_{y\in S} \inpi{y}{\many{y}',q}.\outpi{z}{q})$ for some $i\in I$ (with $Q' \wred Q_i$)
			\item $\respi{\many{x},q_1}(Q'' | \outpi{x_1}{\many{y},q_1} | \sum_{y\in S} \inpi{y}{\many{y}',q}.\outpi{z}{q})$
			\item $\respi{\many x,q}(Q'' | \outpi{z}{q})$
		\end{itemize}
		The latter two are clearly not barbed bisimilar to $A'$.
		For the first case, we have that $S'_I\subseteq S$ and
		$E^{\many{x_i}}_\sigma[Q_i] \wbarb{\out{z_i}}$
		for some $z_i \in S'_I$, so $B'\wbarb{\out{z}}$.
		Thus, $B'$ is not barbed bisimilar to $A'$, because $A'$ is singular
		so we have not $A'\wbarb{\out{z}}$.
		
		\item when $\mu = \earlyinpi{x'}{\many{y}',p_0}$, then $\sigma = \Tag{q_0}{\II},\sigma''$ for some $q_0,\sigma''$ and $\sigma' =
                  \Tag{p_0}{\OO},\sigma$.

                  Given a fresh name $z'$,  we define 
		\begin{align*}
			R \defi & \respi{p_0}(\outpi{x'}{\many{y}',p_0} |  \inpi{p_0}{\many{y}'}.\outpi{x_0}{\many{y}',q_0})\\
			& | \sum_{y\in S'_I} \inpi{y}{\many{y}',q}.\outpi{z'}{q}
		\end{align*}
		
		We note $R'$ the process obtained by applying the law of Lemma~\ref{l:gl} to 
		$\outpi{x'}{\many{y}',p_0}$ in $R$. (Thus $R \bcong{\Tag{q_0}{\OO}} R'$)
		
		We then have to show that for any $z,\many{x},S, Q'$ with \linebreak
		$Q \wtrans{} Q'$,
		we have $A\not\bbis{\Tag{q_n}{\OO}} B$ where
		
		\begin{mathpar}
			A\defi \respi{\many x, q_0}(E^{\many{x_i}}_\sigma[P] | \inpi{z}{q_0}.R | \outpi{z}{q_0})
			\and
			B\defi \respi{\many x, q_0}(E^{\many{x_i}}_\sigma[Q'] | \inpi{z}{q_0}.R' | \outpi{z}{q_0})
		\end{mathpar}
		
		We reason by contradiction. We have \\
		$A \red\red \respi{\many x}(E^{x_0,\many{x_i}}_{\sigma'}[P'] |
		\sum_{y\in S'_I} \inpi{y}{\many{y}',q}.\outpi{z'}{q}) \defi A'$ which is 
		also singular.
		Thus there should exist $B'$ such that $B \wred B'$ and $A' \bbis{\Tag{q_n}{\OO}} B'$.

                We observe that $B'$ can be of 3 forms:
		\begin{itemize}
			\item $\respi{\many x}(E^{x_0,\many{x_i}}_{\sigma'}[Q_i] | \sum_{y\in S'_I}
			\inpi{y}{\many{y}',q}.\outpi{z'}{q})$ for some $i\in I$ \\
			(with
			$Q'|\respi{r_0}(\outpi{x'}{\many{y}',r_0} |
			\inpi{r_0}{\many{y}'}.\outpi{p_0}{\many{y}'})
		\wred Q_i$)
			\item $\respi{\many x}(Q'' | \inpi{z}{q_0}.R' | \outpi{z}{q_0})$
			\item $\respi{\many x}(Q'' | \outpi{z'}{q_0})$
		\end{itemize}
		The latter two are clearly not barbed bisimilar to $A'$.
		For the first case, either $E^{x_0,\many{x_i}}_{\sigma'}[Q_i] \wbarb{\out{z_i}}$ for some $z_i \in S'_I$, so we have that $B'\wbarb{\out{z'}}$.
		Thus, $B'$ is not barbed bisimilar to $A'$, because $A'$ is singular
		so we have not $A'\wbarb{\out{z'}}$.
	\end{itemize}
	
	\textbf{Second case: if $P'$ is not singular.}

        We examine the different possibilities for the transition along $\mu$.
	\begin{itemize}
		\item $\mu = \earlyinpi{x'}{\many{y}',p_0}$, then $\xi = \Tag{q_0}{\II}$ and
		$\sigma' = \Tag{p_0}{\OO},\sigma$. 
		Given fresh names $(z_i)_{i\in I}$, we define 
		\begin{align*}
			R \defi & \respi{p_0}
			(\outpi{x'}{\many{y}',p_0} | \inpi{p_0}{\many{y}'}.\outpi{x_0}{\many{y}',q_0})\\
			& | \sum_{i\in I}\tau.(R_i + \sum_{y\in S'_I}\inpi{y}{\many{y}',q}.\outpi{z_i}{q})
		\end{align*}
		
		and call $R'$ the process obtained by applying the law of Lemma~\ref{l:gl} to $\outpi{x_0}{\many{y}',q_0}$ in $R$. (Thus $R \bcong{\Tag{q_0}{\OO}} R'$)

		We then have to show that for any $z,\many{x},S,Q'$ with \linebreak $Q \wred Q'$, we have
		$A \not\bbis{\Tag{q_n}{\OO}} B$ where
		
		\begin{mathpar}
			A \defi \respi{\many{x},q_0}(E^{\many{x_i}}_\sigma[P] | \inpi{z}{q_0}.R | \outpi{z}{q_0})
			\and
			B \defi \respi{\many{x},q_0}(E^{\many{x_i}}_\sigma[Q'] | \inpi{z}{q_0}.R' | \outpi{z}{q_0})
		\end{mathpar}
		
		We reason by contradiction.
		We have $A \red\red A'$ with
		$$A'\defi \respi{\many{x}}(E^{x_0,\many{x_i}}_{\sigma'}[P'] | \sum_{i\in I}\tau.(R_i + \sum_{y\in S'_I}\inpi{y}{\many{y}',q}.\outpi{z_i}{q}))$$
		
		Thus, there exists $B'$ such that $B \wred B'$ and \linebreak $A' \bbis{\Tag{q_n}{\OO}} B'$.
		As $P'$ is not singular, $E^{x_0,\many{x_i}}_{\sigma'}[P']\wbarb{\out{z'}}$ for some $z'\in S'_I$,
		so $A'\wbarb{\out{z_i}}$ for all $z_i$.
		This implies that $B'$ is of the form $$\respi{x}(E^{x_0,\many{x_i}}_{\sigma'}[Q_i] | \sum_{i\in I}\tau.(R_i + \sum_{y\in S'_I}\inpi{y}{\many{y}',q}.\outpi{z_i}{q}))$$
		
		with $Q'|\respi{r_0}(\outpi{x'}{\many{y}',r_0} |
		\inpi{r_0}{\many{y}'}.\outpi{p_0}{\many{y}'})
		\wred Q_i$.

		We conclude this case by reasoning like in the proof
                of Theorem~\ref{t:completeness}.
                \iflong
                \engue{C'est-à-dire qu'on regarde la clause qui est vraie pour $(P',Q_i)$. Je triche juste sur le fait qu'il peut y avoir une transition en trop qu'on annule avec le Lemma~\ref{l:lin}}
		\fi
                
		\item $\mu = \respi{\many{y}^2}\outpi{p_1}{\many{y}'}$. We note $\many{y}^2 = y^2_1,
		\dots,y^2_n$ and $\many{y}' = \many{y}^2, y^1_1,\dots,y^1_m$. Given fresh 
		names $z',z'',(z_i)_{i\in I}$, we define 
		\begin{align*}
			\hspace{-1em}R & \defi \inpi{x_1}{\many{x}^2,\many{x}^1,q_1}.(\outpi{z'}{q_1} | R_1)\quad\mbox{with}\\
			\hspace{-1em}R_1 & \defi \sum_{j\leq n}\sum_{y''\in\ofnames{P,Q}}[x^2_j = y'']\inpi{z'}{q}.\outpi{z''}{q}\\
			& +
			\sum_{i\in I}[x^1_1 = y^1_1]\dots[x^1_m = y^1_m]\inpi{z}{q}.(\outpi{z_i}{q} | 
			\inpi{z_i}{q'}.R_i)			
		\end{align*}
		
			We then have to show that for any $z,\many{x},S,Q'$ with \linebreak $Q \wred Q'$, we have
		$A \not\bbis{\Tag{q_n}{\OO}} B$ where
		
		\begin{mathpar}
			A \defi \respi{\many{x}}(E^{\many{x_i}}_\sigma[P] | R + \sum_{y\in S} \inpi{y}{\many{y}',q}.\outpi{z}{q}) 
			\and
			B \defi \respi{\many{x}}(E^{\many{x_i}}_\sigma[Q'] | R + \sum_{y\in S} \inpi{y}{\many{y}',q}.\outpi{z}{q})
		\end{mathpar}
	
		We reason by contradiction. We let $\many{x_i}'$ stand
                for $\many{x_i} = x_1,\many{x_i}'$.
		We have $A \wbarb{z}$ and $A \red \red A'$ with 
		$$A' \defi \respi{\many{x},\many{y}^2, q_1}(E^{\many{x_i}'}_{\sigma'}[P'] | R_1).$$
		
		So there should exist $B'$ such that $B \wred B'$ and $A' \bbis{\Tag{q_n}{\OO}} B'$.
		
		As $A'\wbarb{z'}$ and $A'\wbarb{z_i}$ for all $i\in I$, but not $A'\wbarb{z''}$ nor $A'\wbarb{z}$, we have
		that $B'$ can only be of the form $\respi{\many{x},\many{y}^2, q_1}(E^{\many{x_i}'}_{\sigma'}[Q_i] | R_0)$ for some $i\in I$ (with \linebreak$Q' \wtrans{\hat\mu} Q_i$)
		
		We conclude this case by reasoning like in the proof
                of Theorem~\ref{t:completeness}.
\iflong \engue{C'est-à-dire qu'on regarde la clause qui est vraie pour $(P',Q_i)$. Je triche juste sur le fait qu'il peut y avoir une transition en trop qu'on annule avec le Lemma~\ref{l:lin}}		\fi

		\item $\mu = \tau$ and $\xi = \emptyset$. Given fresh names $(z_i)_{i\in I}$,
		we define $R \defi \sum_{i\in I} \tau.(R_i + \sum_{y\in S'_I} \inpi{y}{\many{y}',q}.
		\outpi{z_i}{q})$, and we can conclude.
		
		\item $\mu = \tau$ and $\xi = \Tag{p}{\II}$. Given fresh names $(z_i)_{i\in I}$,
		we define $R \defi \sum_{i\in I} \tau.(\outpi{z_i}{p} | \inpi{z_i}{p}.R_i)$,
		and we can conclude.
		
		\item $\mu = \earlyinpi{p}{\many{y}'}$. Given fresh names $(z_i)_{i\in I}$, 
		we define $$R \defi \outpi{p}{\many{y}'}
		| \sum_{i\in I}\tau.(R_i + \sum_{y\in S'_I}\inpi{y}{\many{y}',q}.\outpi{z_i}{q})$$
\iflong		 \engue{Même triche à la fin avec le Lemma~\ref{l:lin}}\fi
		and we can conclude.

		\item $\mu = \respi{\many{y}^2,q}\outpi{x'}{\many{y}',q}$. We note $\many{y}^2 = y^2_1,
		\dots,y^2_n$ and $\many{y}' = \many{y}^2, y^1_1,\dots,y^1_m$. Given fresh 
		names $z',z'',(z_i)_{i\in I}$, we define
		\begin{align*}
			\hspace{-1em}R &\defi  \inpi{x'}{\many{x}^2,\many{x}^1,q_1}.(\outpi{z'}{q_1} | R_1)\\
			\hspace{-1em}R_1 &\defi   \sum_{j\leq n}\sum_{y''\in\ofnames{P,Q}}[x^2_j = y'']\inpi{z'}{q}.\outpi{z''}{q}\\
			& +
			\sum_{i\in I}[x^1_1 = y^1_1]\dots[x^1_m = y^1_m]\inpi{z}{q}.(\outpi{z_i}{q} | 
			\inpi{z_i}{q'}.R_i)			
		\end{align*}
		We conclude this case by reasoning like in the case for $\mu = \respi{\many{y}^2}\outpi{p_1}{\many{y}'}$.
	\end{itemize}
\end{proof}

Once this proposition is proved, we reason as in the case of
Sequential Bisimilarity to deduce completeness.

\section{Examples with WB Bisimulation (Section~\ref{s:awkward})}\label{a:awkward}

\subsection{A Simplified Example}

We discuss below a simplified example, which exposes the main 
difficulties that arise when studying the well-bracketed state change example. The primary simplification consists in
using single-use functions, i.e., without replication. 
As a consequence,  the two calls to the external function
 are split into two separate functions. 
In  ML, this corresponds to the terms $N$ and $L $ below:
\begin{center}
\begin{tabular}{rcl}
$N$ & $ \defi$ &  \smalltexttt{let x = ref 0 in} $(N_1, N_2)$ \\
$N_1$ & $ \defi$ &  \smalltexttt{fun f -> x := 1; f (); !x} \\
$N_2$ & $ \defi$ &  \smalltexttt{fun f -> x := 0; f (); x := 1}
\\ \ \\ 
$L$ & $ \defi$ &   $(L_1, L_2)$ \\
$L_1$ & $ \defi$ &\smalltexttt{fun f -> f (); 1} \\
$L_2$ & $ \defi$ &\smalltexttt{fun f -> f (); ()}
\end{tabular}
 \end{center}
with the constraint that each term may only be called once. 
(Such a simplification would not work on the well-bracketed state
change example, as $M_1$ and $M_2$
become equivalent even dropping well-bracketing  if used at most once.)

Intuitively, 
\begin{center}
\smalltexttt{let}  $(h_1, h_2) = N$  \smalltexttt{in fun f -> }
$h_2$  \smalltexttt{f} ; $h_1$ \smalltexttt{f}
\end{center}
 is  similar to a single use of
 $M_1$, 
and the same for $L$ and $M_2$.  

The \Api translation of the above terms is as follows: 
	\begin{mathpar}
          \begin{array}{rcl}
\psapp{\enco{N}}{p_0} & \defi & 
                                \respi{\ell,x,y}(\outpi{\ell}{0} | \outpi{p_0}{x,y} | P_1 | P_2)\\
P_1                      & \defi &\inpi{x}{z, p}.\rwriteL{1}.\respi{p'}(\outpi{z}{\star, p'} | p'.\rreadL{n}.\outpi{p}{n})
		\\
P_2                      & \defi& \inpi{y}{z,q}.\rwriteL{0}.\respi{q'}(\outpi{z}{\star,q'} | q'.\rwriteL{1}.\out{q})
\\
            \psapp{\enco{L}}{p_0} & \defi &  \respi{x,y}(\outpi{p_0}{x,y}
                                            | Q_1 | Q_2)\\
Q_1 
 &\defi &\inpi{x}{z, p}.\respi{p'}(\outpi{z}{\star, p'} | p'.\outpi{p}{1})
		\\
            Q_2 
                      &\defi& \inpi{y}{z, q}.\respi{q'}(\outpi{z}{\star, q'} | q'.\out{q})
          \end{array}
	\end{mathpar}
Indeed, the translation of an ML program is parametrised upon a
continuation name (here $p_0$).

We introduce a slight abuse of notation, and write $\psapp{\enco
  N}{p_0}\strans{\respi{x,y}\outpi{p_0}{x,y}}
\enco{N}$, and similarly for notation $\enco L$.

Below we show that ${\enco N}$ and ${\enco L}$  are equivalent.
First however we note that well-bracketing is necessary for this. 
Sequentiality alone is not sufficient: under  the type system for sequentiality
we can observe the following trace from $\enco N$:
	$$
\enco N
        \strans{\earlyinpi{x}{\star,p}}
	\wtrans{\respi{p'}\outpi{z}{\star,p'}}\strans{\earlyinpi{y}{\star,q}}
	\wtrans{\outpi{z}{\star, q'}}\strans{p'}\wtrans{\outpi{p}{0}}$$
	This trace is not well-bracketed: function $x$ is called
        first, but the continuation $p$ is returned before $q$.
Process $\enco L$ may not produce such a trace~--- it may only emit $1$. 
However, the above non-well-bracketed trace is the only trace from 
$\enco N$ that may produce $0$. 

We prove  $\enco N  
 \biswb^{\emptyset} \enco L$ by defining a relation, and relying on up-to techniques for
 WB-bisimulation.

For that, we
use  the abbreviations 
\[ 
\begin{array}{rclrcl}
         P_1' & \defi& p'.\rreadL{n}.\outpi{p}{n} ,
& 
        Q_1'  & \defi& p'.\outpi{p}{1} , \\
        P_2'  & \defi& q'.\rwriteL{1}.\out{q}&
        Q_2'  & \defi& q'.\out{q}
\end{array} 
\]
This allows us to define the following relation, called $\R$:
$$\arraycolsep=1.4pt\begin{array}{l l l}
	\{ (\emptyset,&\respi{\ell}(\outpi{\ell}{0}|P_1|P_2),& Q_1|Q_2),\\
	((\Tag{p'}{\II},\Tag{p}{\OO}),& \respi{\ell}(\outpi{\ell}{1} | P_1' | P_2),&
	Q_1' | Q_2),\\
	((\Tag{q'}{\II},\Tag{q}{\OO}),& \respi{\ell}(\outpi{\ell}{0} | P_1| P_2'),&
	Q_1 | Q_2'),\\
	((\Tag{q'}{\II},\Tag{q}{\OO},\Tag{p'}{\II},\Tag{p}{\OO}),& 
	\respi{\ell}(\outpi{\ell}{0} | P_1'| P_2'),&
	Q_1' | Q_2'),\\
	(\emptyset,& \respi{\ell}(\outpi{\ell}{1}|P_2),& Q_2),\\
	(\emptyset,& \respi{\ell}(\outpi{\ell}{1}|P_1),& Q_1),\\
	((\Tag{p'}{\II},\Tag{p}{\OO},\Tag{q'}{\II},\Tag{q}{\OO}),& 
	\respi{\ell}(\outpi{\ell}{1} | P_1'| P_2'),&
	Q_1' | Q_2'),\\
	((\Tag{p'}{\II},\Tag{p}{\OO}),& \respi{\ell}(\outpi{\ell}{1} | P_1'),& Q_1'),\\
	((\Tag{q'}{\II},\Tag{q}{\OO}),& \respi{\ell}(\outpi{\ell}{0} | P_2'),& Q_2'),\\
	((\Tag{q'}{\II},\Tag{q}{\OO}),& \respi{\ell}(\outpi{\ell}{1} | P_2'),& Q_2'),\\
	(\emptyset,& \respi{\ell}\outpi{\ell}{1},& \nil)\}
\end{array}$$

$\R$ is
 a wb-bisimulation up-to deterministic reduction and up-to static
 context.

\iflong
\ds{above: it would be useful to comment on what are the delicate pairs above; ie the
  pairs in which the constraints on well-bracketing matter}\fi

\noindent\textbf{Without up-to techniques.}

	We define some additional relations, which allow us to
        show the improvement brought by up-to techniques.

	For this, we introduce the following notations for processes
        related to $P_1$ and $Q_1$ respectively:
	\begin{mathpar}
		P_1^a \defi \rwriteL{1}.P_1^b
		\and
		P_1^b \defi \respi{p'}(\outpi{z}{\star,p'} | P_1^c)
		\and
		P_1^c \defi p'.P_1^d
		\and
		P_1^d \defi \rreadL{n}.\outpi{p}{n}
		\\
		Q_1^c \defi p'.\outpi{p}{1}
		\and
		Q_1^b \defi \respi{p'}(\outpi{z}{\star,p'} | Q_1^c)
	\end{mathpar} 
	and similarly for processes $P_2$ and $Q_2$.

	We can remark that $P_i^c$ is the same as $P_i'$ introduced above
        to define $\R$.
        
	\begin{align*}
		\R' \defi\{& (\Tag{p}{\OO}, \respi{\ell}(\outpi{\ell}{0} | P_1^a | P_2), 
		Q_1^a | Q_2),\\
		& (\Tag{p}{\OO}, \respi{\ell}(\outpi{\ell}{0} | P_1 | P_2^a), Q_1 | Q_2^b),\\
		& ((\Tag{q}{\OO},\Tag{p'}{\II},\Tag{p}{\OO}), \respi{\ell}(\outpi{\ell}{1} | P_1^c
		| P_2^a), Q_1^c | Q_2^b)\\
		& (\Tag{p}{\OO}, \respi{\ell}(\outpi{\ell}{1} | P_1^d | P_2), \outpi{p}{1} | Q_2)\\
		& (\Tag{q}{\OO}, \respi{\ell}(\outpi{\ell}{0} | P_1 | P_2^d), Q_1 | \out{q})\\
		& ((\Tag{p}{\OO},\Tag{q'}{\II},\Tag{q}{\OO}), \respi{\ell}(\outpi{\ell}{0} | P_1^a
		| P_2^c), Q_1^b | Q_2^c)\\
		& ((\Tag{q}{\OO},\Tag{p'}{\II},\Tag{p}{\OO}), \respi{\ell}(\outpi{\ell}{0} | P_1^c
		| P_2^d), Q_1^c | \out{q})\\
		& (\Tag{q}{\OO}, \respi{\ell}(\outpi{\ell}{1} | P_2^a), Q_2^b)\\
		& (\Tag{p}{\OO}, \respi{\ell}(\outpi{\ell}{1} | P_1^a), Q_1^b)\\
		& ((\Tag{p}{\OO},\Tag{q'}{\II},\Tag{q}{\OO}), \respi{\ell}(\outpi{\ell}{1} | P_1^d
		| P_2^c), \outpi{p}{1} | Q_2^c)\\
		& (\Tag{q}{\OO}, \respi{\ell}(\outpi{\ell}{0} | P_2^d), \out{q})\\
		& (\Tag{q}{\OO}, \respi{\ell}(\outpi{\ell}{1} | P_2^d), \out{q})\\
		& (\Tag{p}{\OO}, \respi{\ell}(\outpi{\ell}{1} | P_1^d), \outpi{p}{1})\\
		\}
	\end{align*}
	
	\begin{align*}
		\!\R'' \defi\{& (\Tag{p}{\OO}, \respi{\ell}(\outpi{\ell}{1} | P_1^b | P_2), 
		Q_1^b | Q_2),\\
		& (\Tag{p}{\OO}, \respi{\ell}(\outpi{\ell}{0} | P_1 | P_2^b), Q_1 | Q_2^b),\\
		& ((\Tag{q}{\OO},\Tag{p'}{\II},\Tag{p}{\OO}), \respi{\ell}(\outpi{\ell}{0} | P_1^c
		| P_2^b), Q_1^c | Q_2^b)\\
		& (\Tag{p}{\OO}, \respi{\ell}(\outpi{\ell}{1} | \outpi{p}{1} | P_2), \outpi{p}{1} | Q_2)\\
		& (\Tag{q}{\OO}, \respi{\ell}(\outpi{\ell}{1} | P_1 | \out{q}), Q_1 | \out{q})\\
		& ((\Tag{p}{\OO},\Tag{q'}{\II},\Tag{q}{\OO}), \respi{\ell}(\outpi{\ell}{1} | P_1^b
		| P_2^c), Q_1^b | Q_2^c)\\
		& ((\Tag{q}{\OO},\Tag{p'}{\II},\Tag{p}{\OO}), \respi{\ell}(\outpi{\ell}{1} | P_1^c
		| \out{q}), Q_1^c | \out{q})\\
		& (\Tag{q}{\OO}, \respi{\ell}(\outpi{\ell}{0} | P_2^b), Q_2^b)\\
		& (\Tag{p}{\OO}, \respi{\ell}(\outpi{\ell}{1} | P_1^b), Q_1^b)\\
		& ((\Tag{p}{\OO},\Tag{q'}{\II},\Tag{q}{\OO}), \respi{\ell}(\outpi{\ell}{1} | 
		\outpi{p}{1} | P_2^c), \outpi{p}{1} | Q_2^c)\\
		& (\Tag{q}{\OO}, \respi{\ell}(\outpi{\ell}{1} | \out{q}), \out{q})\\
		& (\Tag{p}{\OO}, \respi{\ell}(\outpi{\ell}{1} | \outpi{p}{1}), \outpi{p}{1})\\
		\}
	\end{align*}

	We see that $\R$, $\R'$ and $\R''$ are comparable in size.
        We have that:
        \begin{itemize}
        \item $\R\uplus\R'$ is a wb-bisimulation up-to static context.
	
        \item 
          $\R\uplus\R''$ is a wb-bisimulation up-to deterministic
          $\tau$.
	
        \item 
          $\R\uplus\R'\uplus\R''$ is a wb-bisimulation.
        \end{itemize}

\subsection{Well-bracketed state change}

We present the proof for the example described in
Section~\ref{s:awkward}.

We need to introduce some notations in order to reason about the
sub-processes  which are generated by calls to the functions in the
encodings of $M_1$ and $M_2$.

To simplify notations, we assume that names extruded from $\respi{q}$ (resp.
$\respi{r}$) are in the set $\set{q_i \OR i \in I}$ (resp. $\set{r_j
  \OR j \in J}$)
for some set $I$ (resp. $J$). Continuation names received on $x$ will be noted $p_i$
or $p_j$ depending on which prefix they are underneath.

\begin{mathpar}
	P^1_i = q_i.\respi{r}(\outpi{y}{r} | r.\outpi{p_i}{1})
	\and
	P^2_j = r_j.\outpi{p_j}{1}
	\and
	Q_0 = !\inpi{x}{y,p}.\rwriteL{0}.\respi{q}(\outpi{y}{q} | q.\rwriteL{1}.\respi{r}(\outpi{y}{r} | r.\rreadL{n}.\outpi{p}{n}))
	\and
	Q^1_i = q_i.\rwriteL{1}.\respi{r}(\outpi{y}{r} | r.\rreadL{n}.\outpi{p_i}{n})
	\and
	Q^2_j = r_j.\rreadL{n}.\outpi{p_j}{n}
\end{mathpar}

        Given an ordered list  $s$ of the indices in $I\uplus J$, we
        write $\sigma(s)$ for the stack defined
inductively as follows:
\begin{mathpar}
	\sigma(\emptyset) = \emptyset
	\and
	\sigma(i;s) = \Tag{q_i}{\II},\Tag{p_i}{\OO}, \sigma(s)
	\and
	\sigma(j;s) = \Tag{r_j}{\II},\Tag{p_j}{\OO}, \sigma(s)
\end{mathpar}

We can then define the relation:
\begin{align*}
	\R \defi \{& (\emptyset, P, Q),\quad (\emptyset, P, \respi{\ell}(\outpi{\ell}{1} | Q_0)),\\
		&(\sigma(i;s), P|\prod_{i\in I} P^1_i | \prod_{j\in J} P^2_j,\\
		& \respi{\ell}(\outpi{\ell}{0} | Q_0 | \prod_{i\in I} Q^1_i 
		 | \prod_{j\in J} Q^2_j)),\\
		&(\sigma(i;s), P|\prod_{i\in I} P^1_i | \prod_{j\in J} P^2_j,\\
		& \respi{\ell}(\outpi{\ell}{1} | Q_0 | \prod_{i\in I} Q^1_i 
		| \prod_{j\in J} Q^2_j)),\\
		&(\sigma(j;s), P|\prod_{i\in I} P^1_i | \prod_{j\in J} P^2_j,\\
		& \respi{\ell}(\outpi{\ell}{1} | Q_0 | \prod_{i\in I} Q^1_i 
		| \prod_{j\in J} Q^2_j))\\
	\}
\end{align*}

$\R$ is a wb-bisimulation up-to deterministic $\tau$ and up-to static context.

\end{document}